\documentclass[a4paper,fleqn,usenatbib,useAMS]{mnras}

\usepackage{graphicx}	
\usepackage{amsmath}	
\usepackage{amssymb}	
\usepackage{multicol}        
\usepackage{bm}		
\usepackage{pdflscape}	
\usepackage{journals}

\def\N{NGC\,}

\def\green{\textcolor[rgb]{0.00,0.00,0.00}}
\def\red{\textcolor[rgb]{0.00,0.00,0.00}}

\usepackage[T1]{fontenc}
\usepackage{ae,aecompl}

\usepackage{txfonts}

\title[The insight into the dark side]{The insight into the dark side. \\
I. The \green{pitfalls} of the dark halo parameters estimation
}
\author[A. Saburova et al.]{Anna S. Saburova\thanks{E-mail:
saburovaann@gmail.com}, Anastasia V. Kasparova and Ivan Yu. Katkov
\\
Sternberg Astronomical Institute, Moscow M.V. Lomonosov State University, Universitetskij pr., 13,  Moscow, 119234, Russia\\
}

\begin{document}
\label{firstpage}
\pagerange{\pageref{firstpage}--\pageref{lastpage}} \pubyear{2002}
\maketitle

\begin{abstract}

We examined the reliability of estimates of pseudoisothermal, Burkert and NFW dark halo parameters for \green{ the methods based on the mass-modelling of the rotation curves}.
To do it we constructed the $\chi^2$ maps for the grid of the dark \green{matter} halo parameters for a sample 
of 14~disc galaxies with high quality rotation curves from THINGS. \green{We considered two variants of models in which: a) the mass-to-light ratios of disc and bulge were taken as free parameters, b) the mass-to-light ratios were fixed in a narrow range
according to the models of stellar populations.}

To reproduce the possible \green{observational features of the real galaxies} we made tests showing  that the parameters of the three halo types change \green{critically} in the cases of a lack of kinematic data in the central or peripheral areas and for different spatial resolutions.

We showed that due to the degeneracy between the central densities and the radial scales of the dark haloes there are considerable uncertainties of their concentrations estimates. 
Due to this reason it is also impossible to draw any firm conclusion about universality of the dark halo column density based on mass-modelling of even a high quality rotation curve. 
The problem is not solved by fixing the density of baryonic matter.

In contrast, the estimates of dark halo mass within optical radius are much more reliable. 
We demonstrated that one can evaluate successfully the halo mass using the pure best-fitting method without any restrictions on the mass-to-light ratios.

\end{abstract}


\section{Introduction}
The problem of the dark matter (DM) that was discovered firstly by \citet{Zwicky1933} and then found on galactic scales by \citet{Freeman1970} 
plays important role in the current understanding of the evolution and formation processes in galaxies. 
Despite the long-lasting history of the studies of this problem there are still many open questions. 
For example, there is still no consensus on the form of the DM halo density profiles. 
The problem becomes much more complicated by the fact that the DM density profile could be changed due to the interaction 
with baryons \citep[see, e.g.,][]{Maccio2012}. 
The cosmologically motivated cuspy density profile by \citet{nfw1996} fails to reproduce 
the observations of some galaxies including dwarfs and both high and  low surface brightness galaxies~--- hereafter HSBs and LSBs
\citep[see, e.g.,][]{Burkert1995,Swatersetal2003,deBloketal2001,deBloketal2003,Kuzio2008,
Spano2008,DelPopolo2009,KuziodeNaray2011,2011AJ141193O,2011AJ14224O,little_things}. 
The DM halo-to-total mass fraction also remains quite uncertain 
because it depends on the assumptions that are made in the methods of its determination \citep[see, e.g.,][]{Saburova2014}. 
For \green{disc} galaxies the most popular and frequently used method to estimate the dark halo mass and density profile 
is the rotation curve decomposition into dark and luminous components, it is also called the mass-modelling technique.
This task could have several different solutions especially if there is no additional information on the luminous matter density. \green{It became evident in the  beginning of the studies of the dark matter in galaxies \citep[][]{vanAlbada1986} and it was discussed in the subsequent studies until the present day
 \citep[see, f.e.,][]{Bershady2010, Saburova2014}}. Complementary information can play a critical role. 
For instance, if we assume that the mass-to-light ratio of LSB discs is similar to that of HSBs it gives evidences
 that the dark matter dominates by mass even in the innermost regions. In this case one can explore directly the central 
regions of the dark halo radial distribution in LSBs \citep[see, e.g. zero disc models in][]{Kuzio2008}.
On the contrary, if the stellar discs of these galaxies are heavy, then the DM halo-to-total mass fraction is close to that of normal spirals \citep{Fuchs2003,Saburova2011}.

\green{The pure best-fitting mass-modelling of rotation curve without the additional observational data becomes less popular with the appearance of the photometrical data for big samples of galaxies,  especially in the infrared bands like, e.g., Spitzer Survey of Stellar Structure in Galaxies (S4G), which allows one to determine \green{more} reliably the stellar mass.}
\green{This gave hope that one can finally put the valid constraints on the dark halo shapes, and compare masses and concentrations of the haloes with predicted ones to draw conclusions about the evolution of the galaxies.} \green{However, this possibility should be verified.}

In recent years, researchers claim the universality of such composite dark matter characteristic as its column or surface density:
 $\mu_{0D} \equiv\rho_{0} R_{s}$, where $R_{s}$ is the core radius (or the radial scale) of the density profile, and $\rho_{0}$ is its central density.
\citet{KormendyFreeman2004} obtained the DM surface density for a sample of late type galaxies 
and found that it is almost independent on the B-band luminosity. 
\citet{Donato2009} explored a more numerous sample of galaxies with wider range of luminosities.
They utilized the different mass-modelling techniques and found that the DM halo surface density $\mu_{0D}$ is constant for a wide range of luminosities. 
The conclusion of the universality of the DM halo surface density could be extremely important because it could give additional information about the nature of dark matter. Thus \citet{Milgrom2009} concluded that the constancy of the DM surface density is in good agreement with the modified Newtonian dynamics (MOND) which predicted
 a quasi-universal value of $\mu_{0D}$ for objects of all masses and of any internal structure except LSBs \green{(because these systems possess much lower mean accelerations)}. 
The universality of DM column density can also lead to the important conclusions on the evolution of galaxies of different types. 
\citet{KormendyFreeman2016} showed that it could imply that the spiral, Im and dSph galaxies form a sequence of decreasing baryon-to-DM mass fraction with decreasing luminosity and they even found the hints toward the solution of \textit{too big to fail} problem.

However, the universality of the DM surface density was not confirmed in a number of studies \citep[see, e.g.,][]{Boyarsky2010, Napolitano2010, DelPopolo2013, Saburova2014}. 
The studies of \citet{Saburova2014} and \citet{Napolitano2010} show a dichotomy for the galaxies with low and high masses. 
According to these studies dwarf galaxies have a similar DM surface density \citep[see also][giving evidences for the universality of $\mu_{0D}$ in
 dwarf spheroidal galaxies]{Burkert2015} in contrast to high mass systems for which this parameter varies with luminosity and mass. 

In current paper we present a research of the popular mass-modelling \green{of the rotation curve, which is widely used for the disc galaxies.}
Our goal is to test the reliability of the estimates of dark halo parameters using high quality input data. 
Using our technique we can also study the interconnection between the halo parameters and test the universality of the DM halo surface density.

\green{The paper is organized as follows: the used method, the sample and the utilized data are characterized in Sect.~\ref{Method}. The main results and discussion are given in Sect.~\ref{results} and  Sect.~\ref{discussion}. 
Section~\ref{Conclusions} is devoted to the summary.}

\section{The method and the data}
\label{Method}

The concept of universality of the DM surface density gave us idea that the rotation curve of a galaxy could be equally good fitted by several different pairs of values of radial scale and central density of DM halo, and there could be interdependence between these parameters in a galaxy. 
\green{ For models with fixed density of baryonic matter }  \green {this interconnection is mostly due to the halo shape \citep{deBloketal2001}. 
For example for the pseudoisothermal DM halo profile the central density and the core radius are linked through the asymptotic velocity $v^2_{\mathrm{as}} \propto \rho_{0} R_{s}^2 $, which is restricted by the velocity on the flat part of the rotation curve.}  
\green{However, in general case when the stellar density is considered as a free parameter this relation is less evident.}

To study the interconnection between the dark halo parameters and their uncertainty we developed a technique, the main point of which is similar to that used in some studies \citep[e.g.,][]{vandenBosch2000, deBloketal2001, Napolitano2014, Corbelli2014}. 
Its basic concept is to calculate the $\chi^2$ for the grid of values of the DM halo parameters. 
The advantage of the current paper is that we study the behavior of $\chi^2$ maps in details for a sample of galaxies with accurate rotation curves and also analyze the connection of the $\chi^2$ map features with the shape, the extension and spatial resolution of the rotation curves. 
\green{This study allows to get an idea about the reliability of output halo parameters simply by looking at the features of the input kinematical data.}

\green{We consider both the photometrical approach\footnote{The recent papers  using the similar method: \citet{THINGS, little_things, Frank2015,Karachentsev2016}.}, in which the stellar surface density is found from the photometry (mostly in infrared band), and the pure best-fitting technique\footnote{This approach was utilized for example by the following papers: \citet{Barnesetal2004, Cheminetal2006, Spano2008, THINGS, CardoneDelPopolo2012, Kasparova2012,Sofue2015,Katz2016}.}, when the mass-to-light ratios of stellar disc and bulge represent the free parameters. }

\subsection{\green{The used equations}}

We consider disc galaxies with rotation curves $v(r)$ that could be decomposed into the contributions of bulge, gas and stellar discs and DM halo: 
\begin{equation}    
v^2(r)=v^2_{bulge}(r)+v^2_{disc}(r)+ v^2_{gas}(r)+v^2_{halo}(r).
\end{equation}  
The contribution of the DM halo is defined by the parameters of its density profile. 
In the current paper we used three \green{halo} types:

(i) The density profile by \citet{Burkert1995}: 
\begin{equation}\label{Burkert}
\rho_{\mathrm{burk}}(r)=\frac{\rho_0 R_s^3}{(r+R_s)(r^2+R_s^2)}. 
\end{equation} 
Here $\rho_{0}$ and $R_{s}$ are the central density and the radial scale of the halo\footnote{Below $R_s$ and $\rho_0$ are different for the various DM density profiles.}.  
The corresponding contribution to the rotation curve: 
\begin{equation}
\begin{split}
(v^2_{halo}(r))_{\mathrm{burk}}&=
6.4G\frac{\rho _{0}R_{s}^{3}}{r}\left[ \ln \left( 1+\frac{r}{R_{s}}\right) - \right.\\
&\left.  -\arctan \left( \frac{r}{R_{s}}\right) +\frac{1}{2}\ln
\left(1+\left( \frac{r}{R_{s}}\right) ^{2}\right) \right], 
\end{split}
\end{equation} 
where $G$ is the gravitational constant.
We calculated \green{the halo masses} within optical radius\footnote{\green{The values of $R_{opt}$ are given in Tables ~\ref{parameters_burk}--\ref{parameters_nfw}. For most galaxies of the sample \cite[except \N2976 and \N4736, for which these values were taken from][]{Leroy2008} we assumed that the optical radius is equal to four disc radial scalelengths.}} $R_{opt}$. \green{In case of Burkert profile, it is given by} the following formula:
\begin{equation}
\begin{split}
(M_{halo})_{\mathrm{burk}} &= 2\pi\rho_0 R_s^3 \left[\ln\left(1+\frac{R_{opt}}{R_s}\right)+ \right.\\
&\left.+\frac{1}{2}\ln\left(1+\left(\frac{R_{opt}}{R_s}\right)^2\right) - \arctan\left(\frac{R_{opt}}{R_s}\right) \right].
\end{split}
\end{equation}

(ii) The pseudoisothermal profile (hereafter, piso):  
\begin{equation}    
\rho_{\mathrm{piso}}(r)=\frac{\rho_{0}}{(1+(r/R_{s})^2)},
\end{equation} 

\begin{equation}
(v^2_{halo}(r))_{\mathrm{piso}} =  4\pi G\rho_{0} R_{s}^2 \Bigl[ 1 -
{{ R_{s}}\over{r}}\arctan \Bigl( {{r}\over{R_{s}}} \Bigr) \Bigr],
\end{equation}
\begin{equation}
(M_{halo})_{\mathrm{piso}} = 4\pi \rho_0 R_s^2\left[R_{opt}-R_s \arctan\left(\frac{R_{opt}}{R_s}\right)\right].
\end{equation}

(iii) The Navarro-Frenk-White profile \citet{nfw1996}  (hereafter, NFW): \begin{equation}    
\rho_{\mathrm{nfw}}(r)=\frac{\rho_{0}}{(r/R_s)(1+(r/R_{s })^2)^{2}},
\end{equation} 
 
\begin{equation}
(v^2_{halo}(r))_{\mathrm{nfw}} =  4\pi G\rho_{0} R_{s}^3/r \Bigl[ \log(1+r/R_s)- \frac{r/R_s}{1+r/R_s}\Bigr],
\end{equation}
\begin{equation}
(M_{halo})_{\mathrm{nfw}} = 4\pi\rho_0R_s^3\left[\ln\left(\frac{R_s+R_{opt}}{R_s}\right)-\frac{R_{opt}}{R_s+R_{opt}}\right].
\end{equation}

For the bulge component we adopted Sersic profile of~surface density \citep{Sersic68}:
\begin{equation}
\label{eq:Sersic}
I_b(r) = 
          (I_0)_b 10^{\left[ - b_n\left( \frac{r}{R_{e}} \right)^{1/n} \right]}. \end{equation}
Here $(I_0)_b$ is the bulge central surface density, $R_e$ is the effective radius containing a half of the luminosity,
$b_n\approx 1.9992 n - 0.3271$ (\citealt{Caon1993})
and $n$ is the Sersic index. 
We used the rotation curve of Sersic bulge determined according to \citet{Noordermeer2008}.

We calculated the gas and stellar disc contributions to the rotation curve \green{using the method by \citet{Casertano1983}} for a non-parametric density profile. 
The surface density profile of gas, the disc radial scalelength, bulge effective radius and \green{Sersic index} were fixed and taken from observations (see Sect.~\ref{Sample}). For the galaxies with available S4G images the surface density radial profile of stellar disc was determined from a difference between the total observed surface brightness profile and that of the bulge (see below). 

\subsection{\green{The assumptions on mass-to-light ratios: the two models}}
\label{ml}

\green{We considered the two possibilities for mass-to-light ratios of the stellar bulge and disc which were independent parameters during the fitting. In \green{the Models~A} the disc and bulge mass-to-light ratios were considered as free parameters (the best-fitting approach). In \green{the Model~B} we allowed them to vary in a narrow range between 0.45 \citep[obtained using Tully-Fisher relation in a good agreement with the models of stellar population by][]{McGaugh2015} and 0.6 \citep{Meidtetal2014} solar units at $3.6\mu \mathrm{m}$\footnote{\green{For \N0925 and \N7331 absented in S4G with photometrical data in V-band we used the ranges of $M/L_V$: $0.54-0.88$ and $1.62-1.9$, correspondingly, according to their total $(B-V)_0$ colors and model relations from \cite{Zibetti2009} for Chabrier IMF and \cite{Belletal2003} for scaled Salpeter IMF. }}. The value of the mass-to-light ratio at $3.6\mu \mathrm{m}$ is almost independent on the color of stellar population \citep{McGaughSchombert2014}, thus it is reasonable to consider it to be nearly constant for all galaxies of the sample. But for three galaxies of our sample (\N2903, \N5055, \N6946) we had to widen the range of possible mass-to-light ratios to 0.3-0.6 in order to avoid unacceptably bad fit of the rotation curve.}

\green{The pure best-fitting model (Model~A) is worth consideration despite it becomes less popular. The reasons are the following. Model A can be regarded as the most general case of the technique of mass-modelling of a rotation curve. If for example some DM halos profiles fail to reproduce the observed rotation curve even in this general case, they will definitely fail when the baryonic surface density is fixed. Another issue: in some cases, the photometrical model gives unsatisfactorily fit to the rotation curve (see the discussion of this problem in Sect. \ref{comp}). }

\green{The results of our analysis of the best-fitting technique could also be useful if one needs to study the samples of galaxies for which there are no homogeneous high quality surface photometry data and no other information, that can be used to obtain the independent estimate of the disc surface density. For some unusual objects, like LSBs or galaxies with low dynamical mass-to-light ratios \citep[see][]{Saburova2015} the photometrical method can be less plausible than the best-fitting modelling, because there is possibility that the model color--$M/L$ relations for standard IMF (e.g. Salpeter) are not valid for them. }

Thus, in order to understand the uncertainty of the dark halo parameters in general one should examine it for both fixed and free baryonic density distribution.

\subsection{The sample and the data}
\label{Sample}

We \green{performed our analysis} on the sample of HSB spiral galaxies with accurate rotation curves from The H{\sc i} Nearby Galaxy Survey \citep[THINGS,][]{THINGS}. 
The high spatial and velocity resolution made these data ideal for this aim. 
To put constraints on the stellar surface densities we choose the 3.6\ $\mu$m data from Spitzer Survey of Stellar Structure in Galaxies \citep[S4G,][]{s4g}. 
Unfortunately, these data were available not for all galaxies of the sample. 
For the galaxies absented in the S4G we used the structural parameters of disc and bulge taken from other literature. 
In Table~\ref{tab_sample} we give the names of the galaxies, the adopted distances and the note on the source of photometric data. 

When the photometric data were available in S4G we firstly constructed the radial surface brightness profiles using ELLIPSE routine of IRAF software (\citealt{iraf}). 
After that we decomposed the profiles into the contribution \green{of Sersic bulge (Eq.~\ref{eq:Sersic}) and exponential disc}: 
\begin{equation}I_d(r)=(I_{d})_{0}\exp(-r/R_d), \end{equation} where $(I_{d})_{0}$ and $R_d$ are the disc central surface brightness and the exponential scalelength, correspondingly.  
Since in some cases the special features on the surface density profile (like knees) can play important role in the decomposition of the rotation curve we did not
 use in our analysis the smoothed radial density profile of exponential disc for the galaxies with available S4G data except \N2976. We came further and subtracted the
 contribution of the bulge from the total surface brightness profile to estimate the non-parametric surface brightness profile of the disc which we used in our analysis
 together with the bulge profile. \green{The calibration of the profiles was performed according to \cite{Querejeta2015}. We also corrected the 3.6\ $\mu$m surface brightness for the inclination following \cite{Graham2001}. The \green{inclination angles} were taken from \citep{THINGS}.}

The parameters of bulges for the galaxies contained in S4G (not corrected for the inclination) are given in Table~\ref{bulges}.  
For \N2976 we neglected the contribution of bulge to the total surface brightness profile following \citet{THINGS}. 

To estimate the contribution of gas discs to the rotation curves we used the H{\sc i} surface density radial profiles obtained by \cite{THINGS} and the molecular gas density profiles from \cite{Leroy2008} (\green{both profiles were corrected for the presence of Helium}). 
 
\subsection{\green{Visualization of the results}}

We calculated the values of $\chi^2$ for a grid of parameters of the DM halo. 
The ${\chi^2}$ minimization was performed using the constrained non-linear Levenberg--Marquardt minimization implemented in the {\sc mpfit} IDL package (by C.~Markwardt, NASA). We believe that the choice of method of finding of $\chi^2$ minimum does not influence the results. 

The resulting $\chi^2$ maps demonstrate the goodness of the rotation curve fitting for given parameters of DM halo. 
Each point of the map corresponds to the concrete values of the mass-to-light ratio of disc and bulge which, in cases of Models~A, were varied without any limits in order to get the best-fitting of the rotation curve, while the dark halo radial scale and central volume density were fixed. 
\green{For Model B it is the same, but  the mass-to-light ratios are varied in a narrow ranges given above.}

\green{In Figs.~\ref{fig1}--\ref{fig4} we show examples of the $\chi^2$ maps for both A and B Models.} 
Left panels correspond to the models with the piso profiles of the dark halo, centre panels~--- to the Burkert  profiles, right panels~--- to the NFW profiles.
The color on the maps denotes the $\chi^2$ value, the darker the color, the lower the $\chi^2$ and the better is the fit. 
The white contours refer to $1\sigma$, $2\sigma$ and $3\sigma$ confidence limits. 
The position of the parameters resulting from decomposition corresponding to the $\chi^2$ minimum is shown by cross in each map. 
\green{Black solid and dashed curves refer to the lines of constant mass of the dark halo inside of optical radius $R_{opt}$ calculated from the parameters of the $\chi^2$ minimum models $M_{halo}\pm15$~per~cent.}
Blue dot-and-dash line corresponds to the constant surface density of dark halo $\log \mu_{0D} =2.15$ taken from \citet{Donato2009} in the case of Burkert    density profile. 
To convert $\mu_{0D}$ into corresponding values of NFW and piso profiles
we used the translation formulas from \cite{Boyarsky}: 
\begin{equation}
\begin{split}
(R_{s})_{\mathrm{piso}}&=0.26(R_{s})_{\mathrm{burk}}, \\
(\rho_{0})_{\mathrm{piso}}&=3.36 (\rho_{0})_{\mathrm{burk}},
\end{split}
\label{eq9}
\end{equation}    
\begin{equation}    
\begin{split}
(R_{s})_{\mathrm{nfw}}&=1.6(R_{s})_{\mathrm{burk}}, \\
(\rho_{0})_{\mathrm{nfw}}&= 0.37 (\rho_{0})_{\mathrm{burk}}.
\end{split}
\label{eq10}
\end{equation}
\green{Red lines (dotted and dash-dot-dot-dotted) show the positions of the constant ratios $M/L$, which were discussed in Sect. \ref{ml}.}

\green{For illustrative purposes we also performed best-fitting of the rotation curves corresponding to the $\chi^2$ minimum (the lower row for each galaxy in Figs.~\ref{fig1}--\ref{fig4}). The} black dots with error bars demonstrate the observed rotation curve \citep{THINGS}, thick red line~--- the total model, thin black line~---  dark halo, blue dashed line~--- stellar disc, dot-and-dashed red line~---  gas disc, dotted line~---  bulge.

We also give in Tables~\ref{parameters_burk}--\ref{parameters_nfw} the parameters corresponding to the minimal values of $\chi^2$ for each type of DM density profile \green{for both A and B Models} together with the errors associated with the range covered by $1\sigma$ confidence limit (each pair of the parameters of DM halo on the map corresponds to the certain values of $\chi^2$, disc mass-to-light ratio and bulge central surface density) and \green{the minimal values of $\chi^2$ divided by the number of degrees of freedom}. 
The listed estimates of disc mass-to-light ratio $M/L$ correspond to 3.6\ $\mu$m for all galaxies except for V-band cases of \N925 and \N7331.

\begin{table}
\begin{center}
\caption{The sample. \label{tab_sample}}
\begin{tabular}{lll}
\hline
\N &Dist., Mpc&Photometry source\\
\hline
\hline
0925	&	9.2	&	\cite{Baggett1998}	\\
2403	&	3.2	&	\cite{THINGS}	\\
2841	&	14.1	&	S4G	\\
2903	&	8.9	&	S4G	\\
2976	&	3.6	&	S4G	\\
3031	&	3.6	&	S4G	\\
3198	&	13.8	&	S4G	\\
3521	&	10.7	&	S4G	\\
3621	&	6.6	&	\cite{THINGS}	\\
4736	&	4.7	&	S4G	\\
5055	&	10.1	&	S4G	\\
6946	&	5.9	&	\cite{THINGS}	\\
7331	&	14.7	&	\cite{Baggett1998}	\\
7793	&	3.9	&	S4G	\\
\hline
\end{tabular}
\end{center}
\end{table}

\begin{table}
\begin{center}
\caption{The structural parameters of bulges for S4G galaxies: \green{effective radius, central surface brightness in 3.6$\mu \mathrm{m}$ and Sersic index.}
\label{bulges}}
\begin{tabular}{llll}
\hline
\N & $R_{e}$ & $	\nu_{0~(3.6\mu \mathrm{m})}$  & $	n	$\\
& arcsec & mag/arcsec$^2$&\\
	\hline					
	\hline	
2841	 & $	9.2 \pm   0.2	$ & $	11.80 \pm   0.03	$ & $	1.7\pm  0.1	$ \\
2903	 & $	6.2 \pm  0.1	$ & $	12.94 \pm   0.01	$ & $	0.8\pm   0.0	$ \\
2976	 & $	-	$ & $	-	$ & $	-	$ \\
3031	 & $	20.3 \pm  0.7	$ & $	11.85 \pm  0.05	$ & $	1.4\pm  0.1	$ \\
3198	 & $	4.3 \pm  0.2	$ & $	13.25 \pm   0.07	$ & $	 1.7 \pm 0.1	$ \\
3521	 & $	11.3 \pm   0.8	$ & $	13.55 \pm   0.08	$ & $	2.1 \pm 0.2	$ \\
4736	 & $	10.86 \pm   0.0	$ & $	14.73 \pm   0.01	$ & $	1.0 \pm   0.0	$ \\
5055	 & $	22.13 \pm  9.6	$ & $	13.45 \pm   0.96	$ & $	2.6 \pm   0.1	$ \\
7793	 & $	 7.7 \pm   0.4 	$ & $	16.07 \pm  0.04	$ & $	2.0 \pm   0.1	$ \\

\hline
\end{tabular}
\end{center}
\end{table}

\begin{figure*} 
\includegraphics[width=15.5cm]{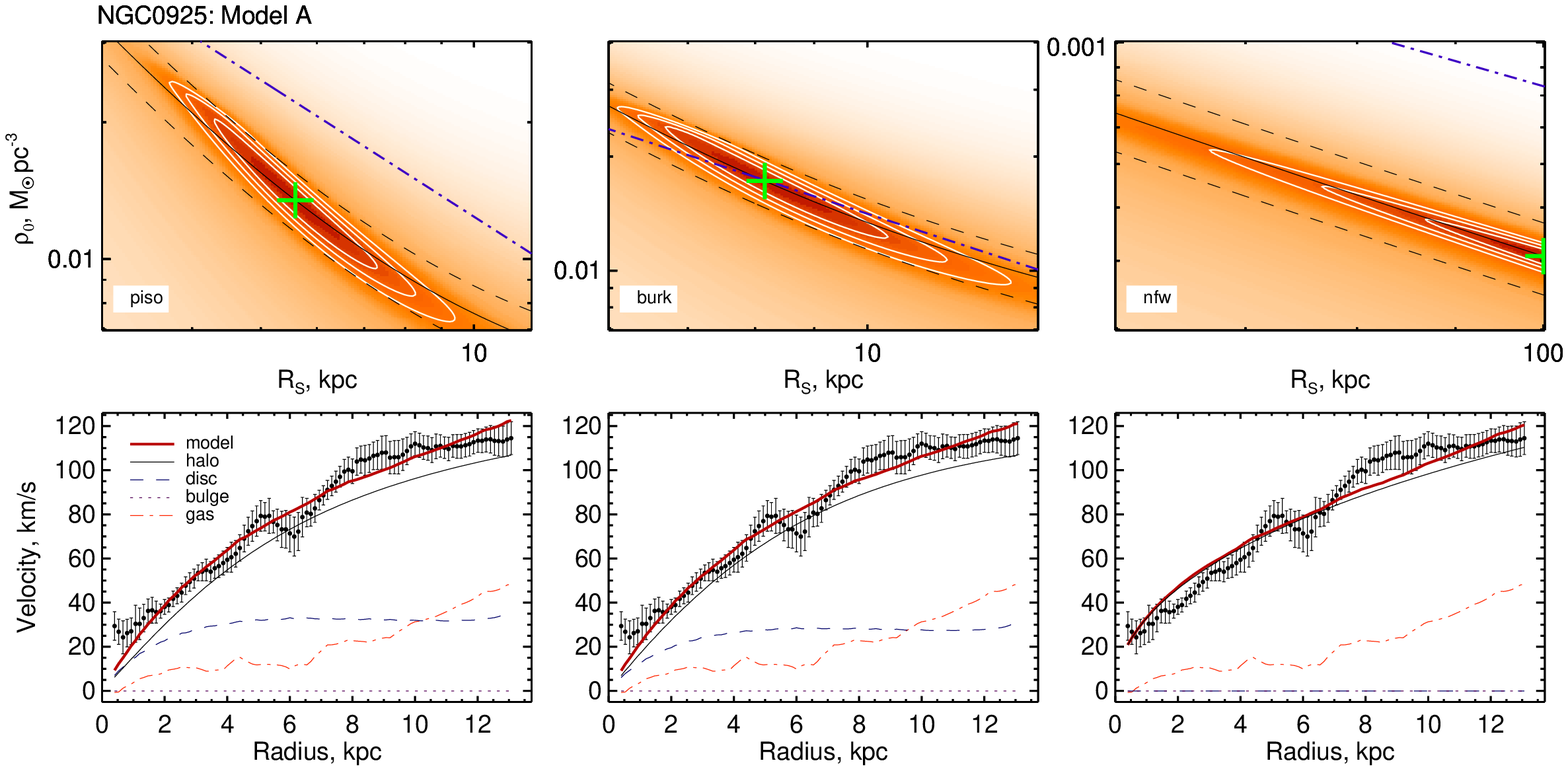}
\includegraphics[width=15.5cm]{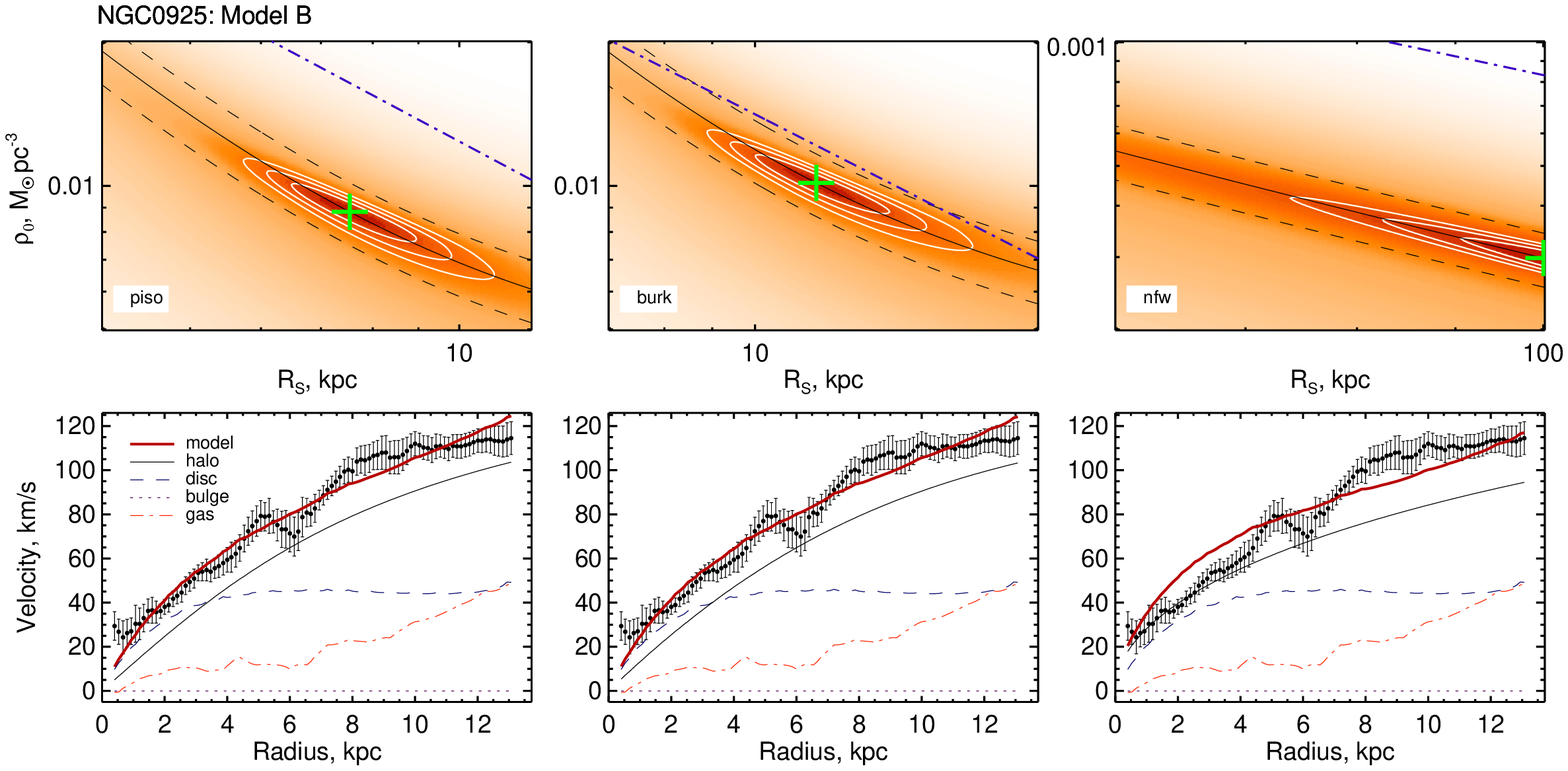}
\caption{
The examples of $\chi^2$ maps for galaxies with increasing rotation curves. 
\green{Two top rows for each galaxy show best-fitting approach (Model~A), two bottom rows for each galaxy are for photometric approach (Model~B).} 
Left panels correspond to the models with the piso profiles of the DM halo, centre panels~--- to the Burkert profiles, right panels~--- to the NFW profiles.
The color in the maps denotes the $\chi^2$ value, the darker the color, the lower the $\chi^2$ and the better is the fit. 
The white contours refer to $1\sigma$, $2\sigma$ and $3\sigma$ confidence limits. 
The position of the parameters corresponding to the $\chi^2$ minimum is shown by green cross in each map. 
Red lines (dotted and dash-dot-dot-dotted) show the positions of the constant ratios $M/L$ of discs (see text). 
Blue dot-and-dash line corresponds to the constant surface density of dark halo. 
Black lines in the $\chi^2$ maps refer to the lines of constant mass of the dark halo inside of optical radius (calculated from the parameters of $\chi^2$ minimum model) $M_{halo}\pm15$~per~cent. 
The lower row for each galaxy gives the best-fitting decomposition \green{corresponding to the $\chi^2$ minimum}, where the black dots with error bars mark the observed rotation curve \citep{THINGS}, thick red line~--- total model, thin black line~--- dark halo, blue dashed line~--- stellar disc, dot-and-dashed red line~--- gas disc, dotted line~--- bulge.}
\label{fig1}
\end{figure*}
\addtocounter{figure}{-1}
\begin{figure*} 
\includegraphics[width=16.5cm]{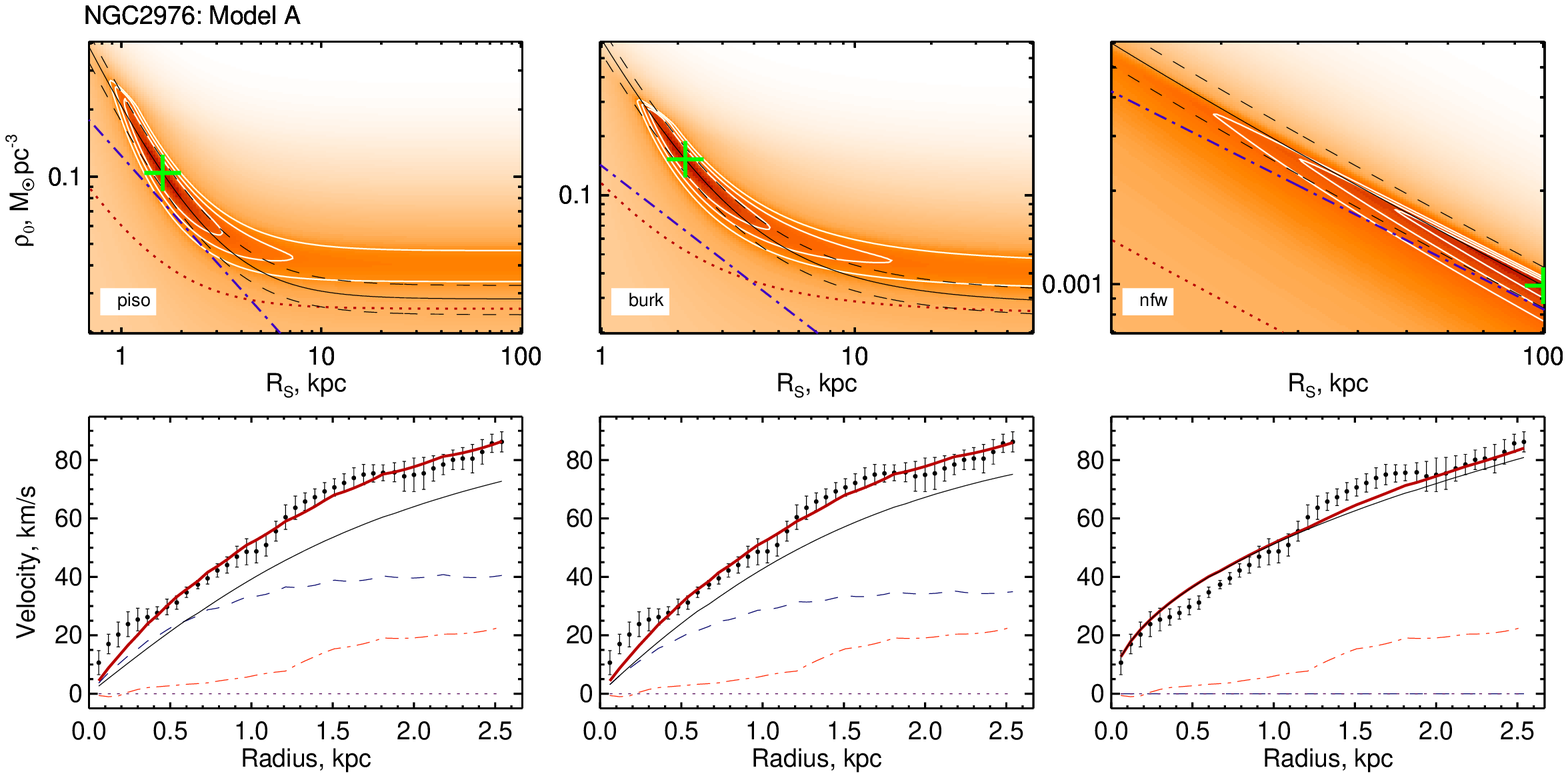}
\includegraphics[width=16.5cm]{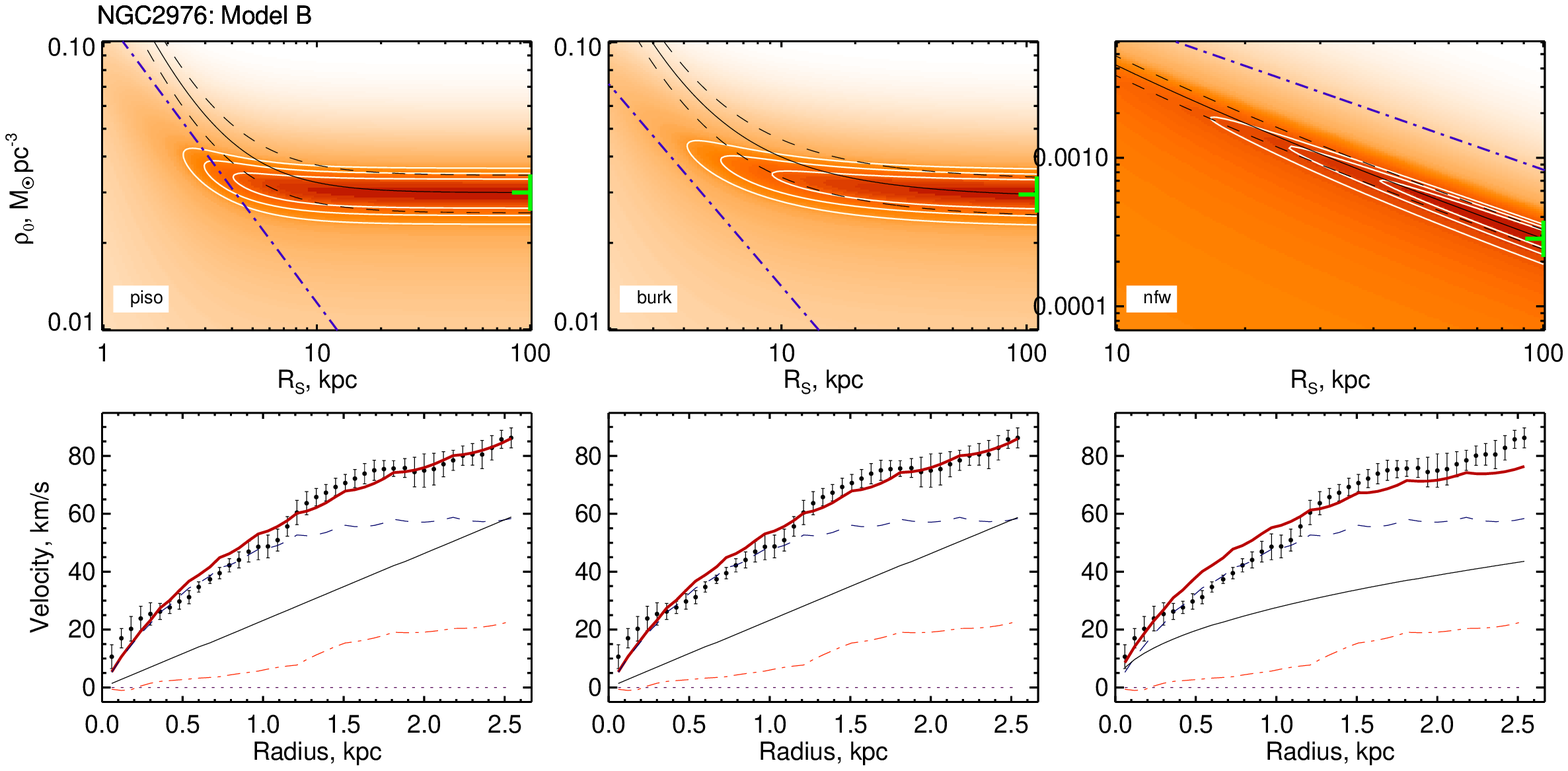}
\caption{(Continued).}
\end{figure*}

\begin{figure*} 
\includegraphics[width=16.5cm]{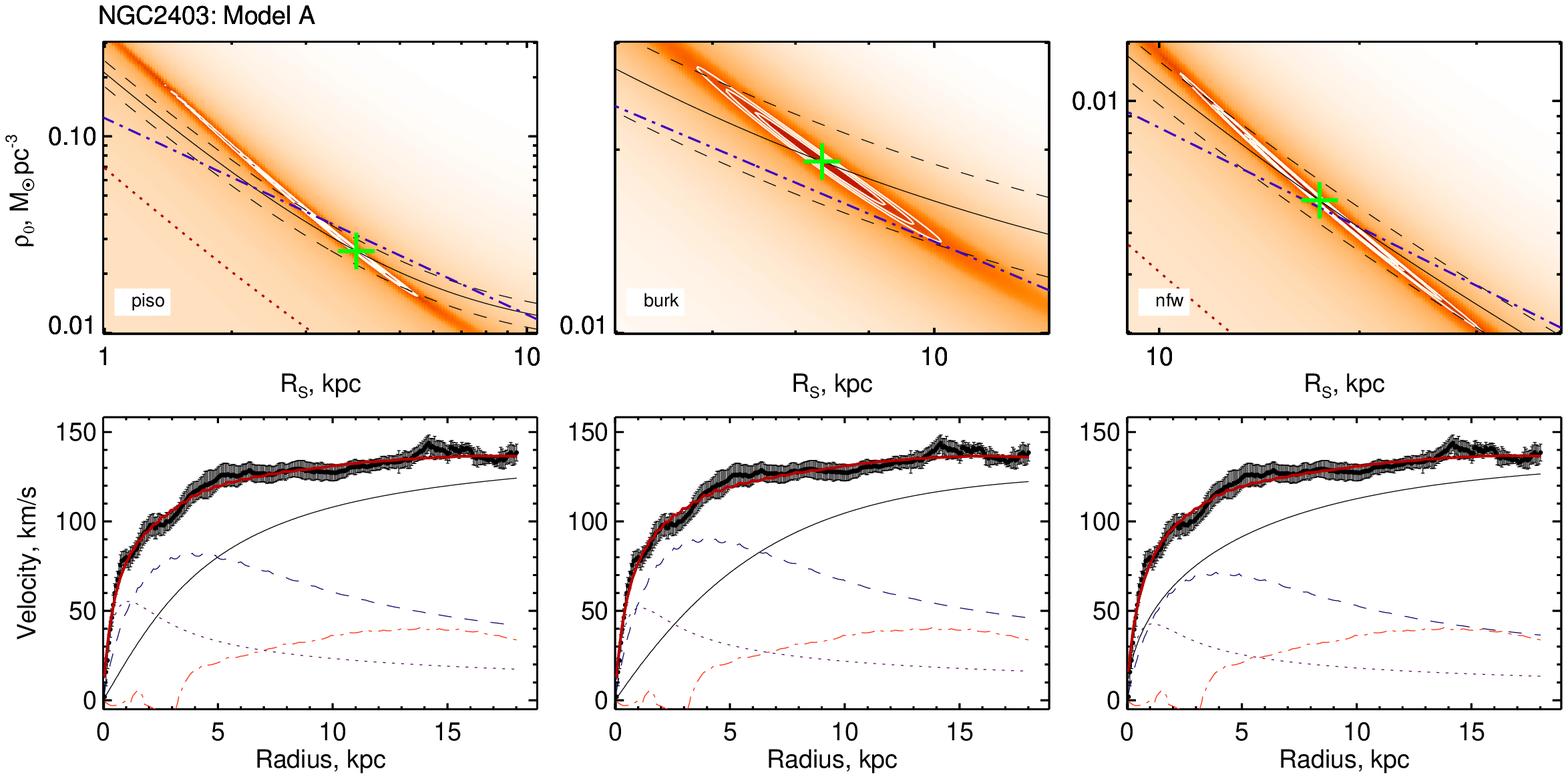}
\includegraphics[width=16.5cm]{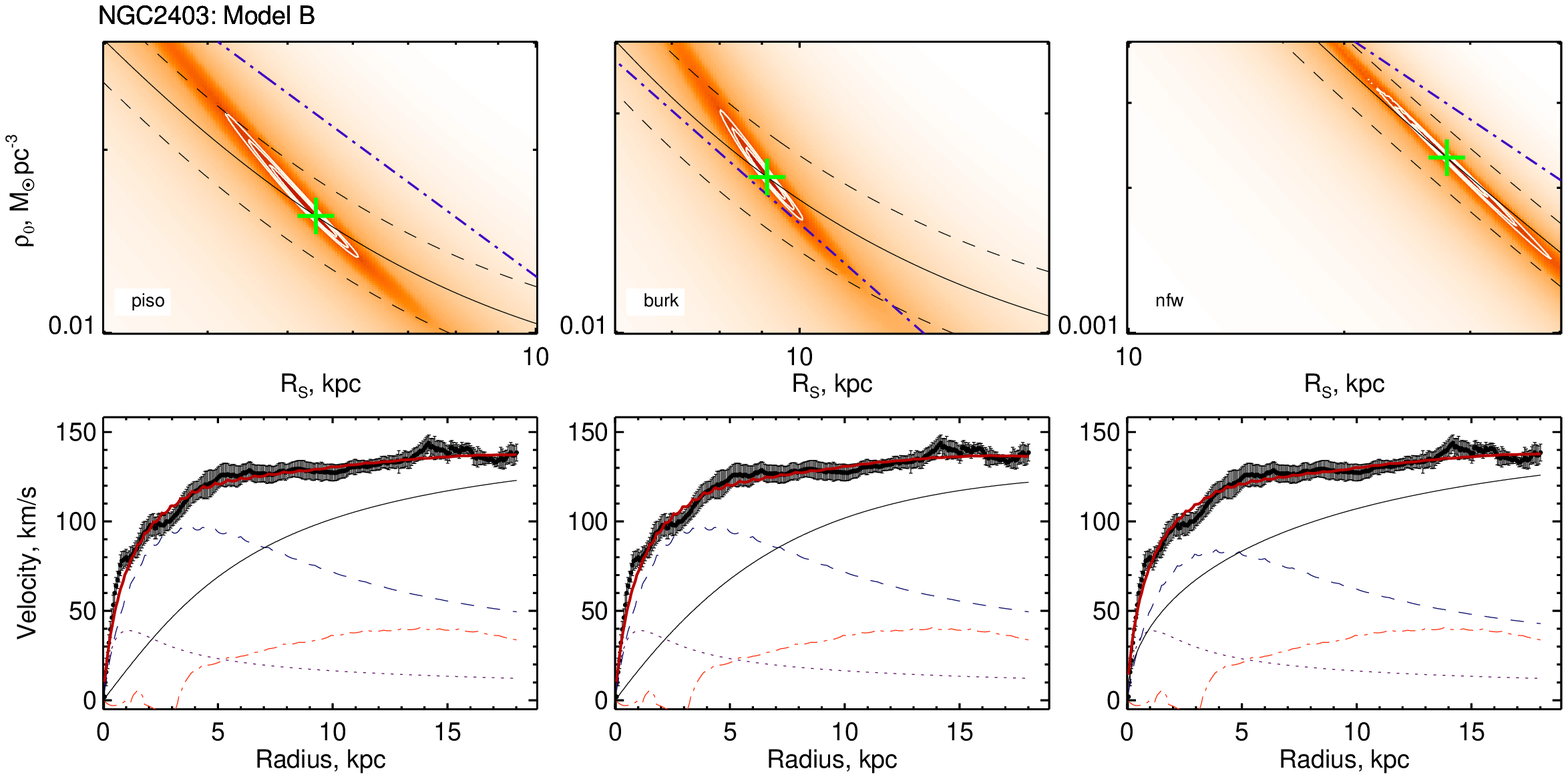}

\caption{The same as in Fig~\ref{fig1}, but for the cases of detailed data in the inner parts of rotation curves. }
\label{fig2}
\end{figure*}
\addtocounter{figure}{-1}
\begin{figure*} 
\includegraphics[width=16.5cm]{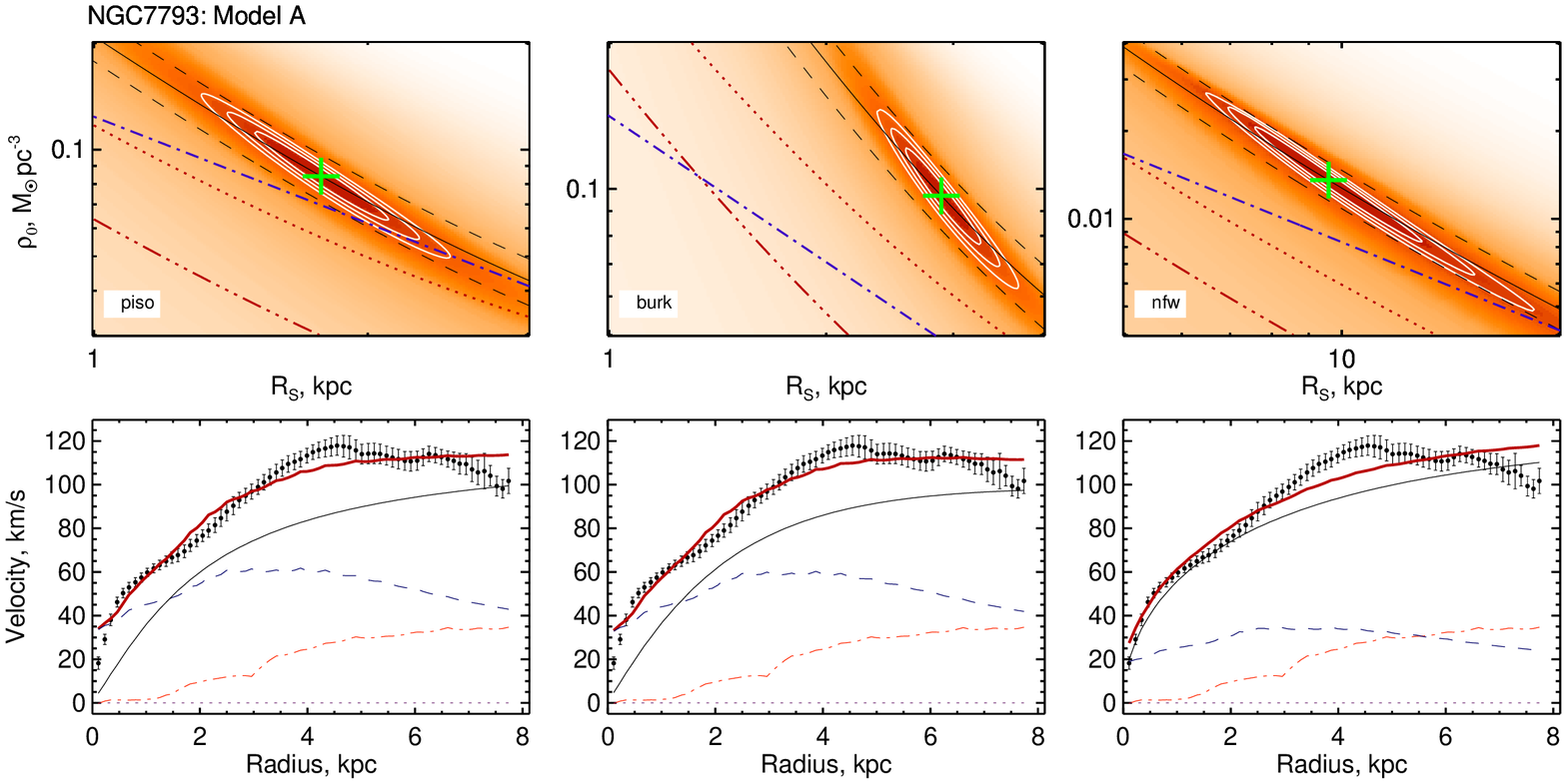}
\includegraphics[width=16.5cm]{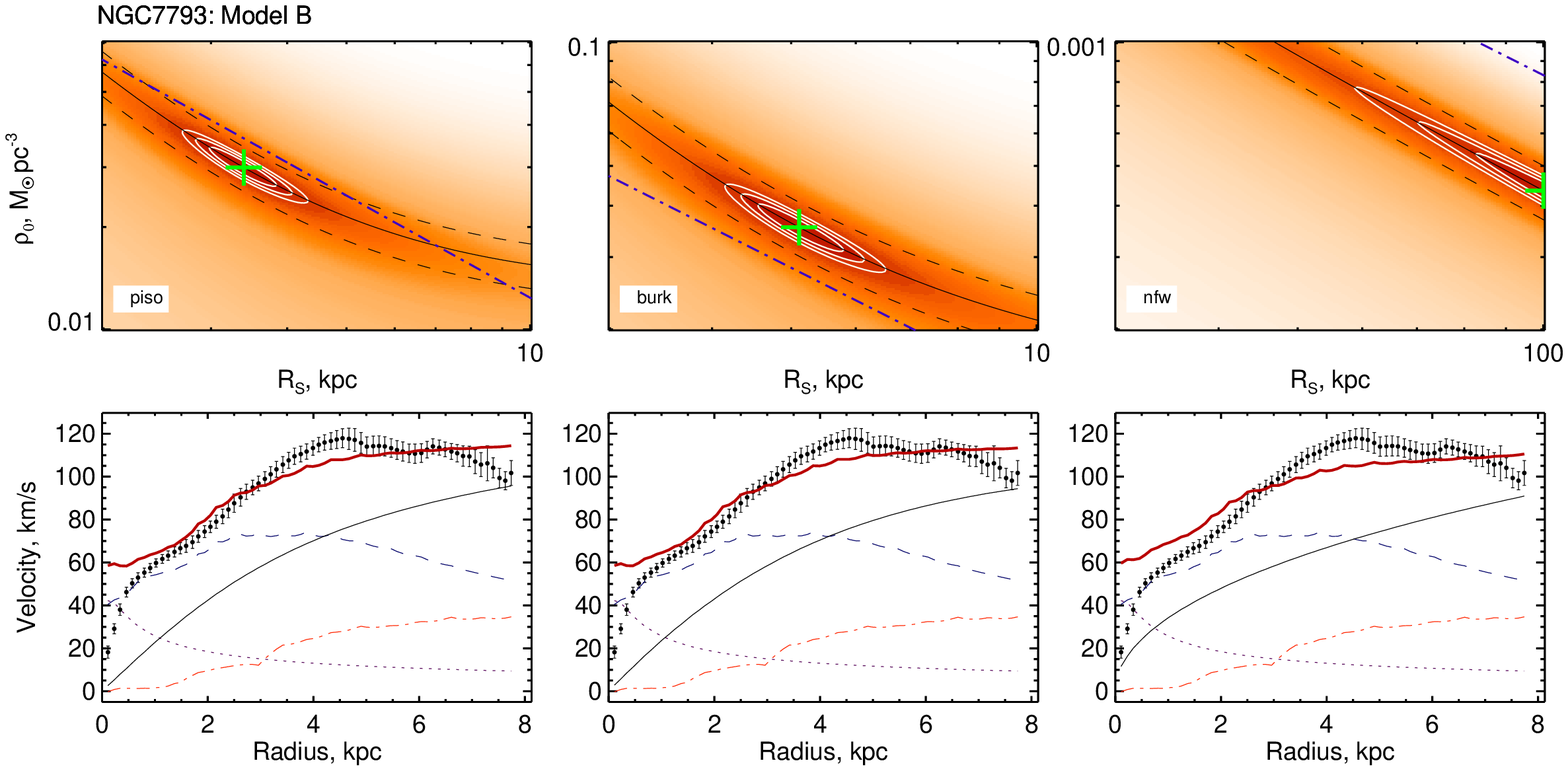}
\caption{(Continued).}
\end{figure*}
\addtocounter{figure}{-1}
\begin{figure*} 
\includegraphics[width=16.5cm]{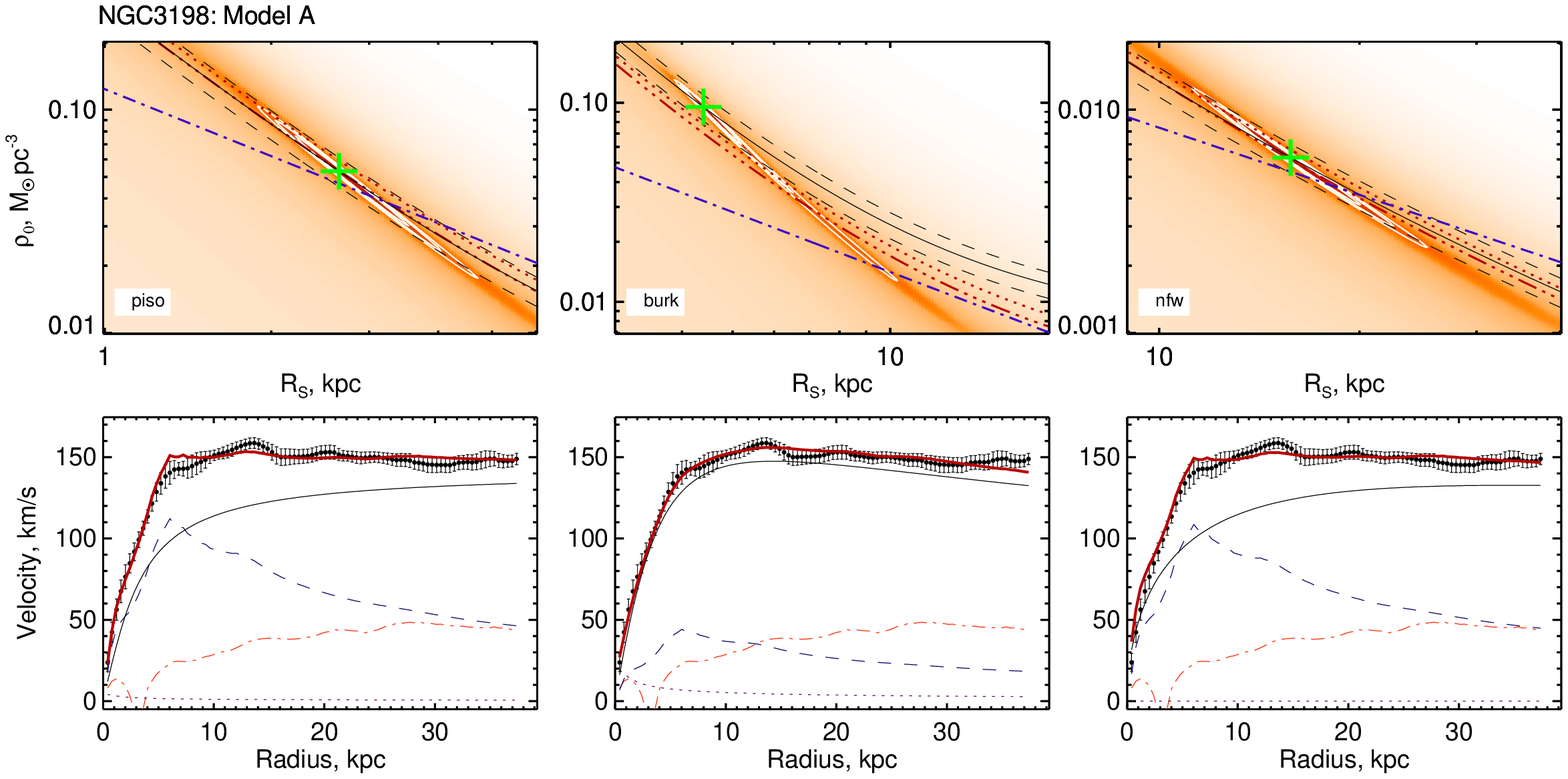}
\includegraphics[width=16.5cm]{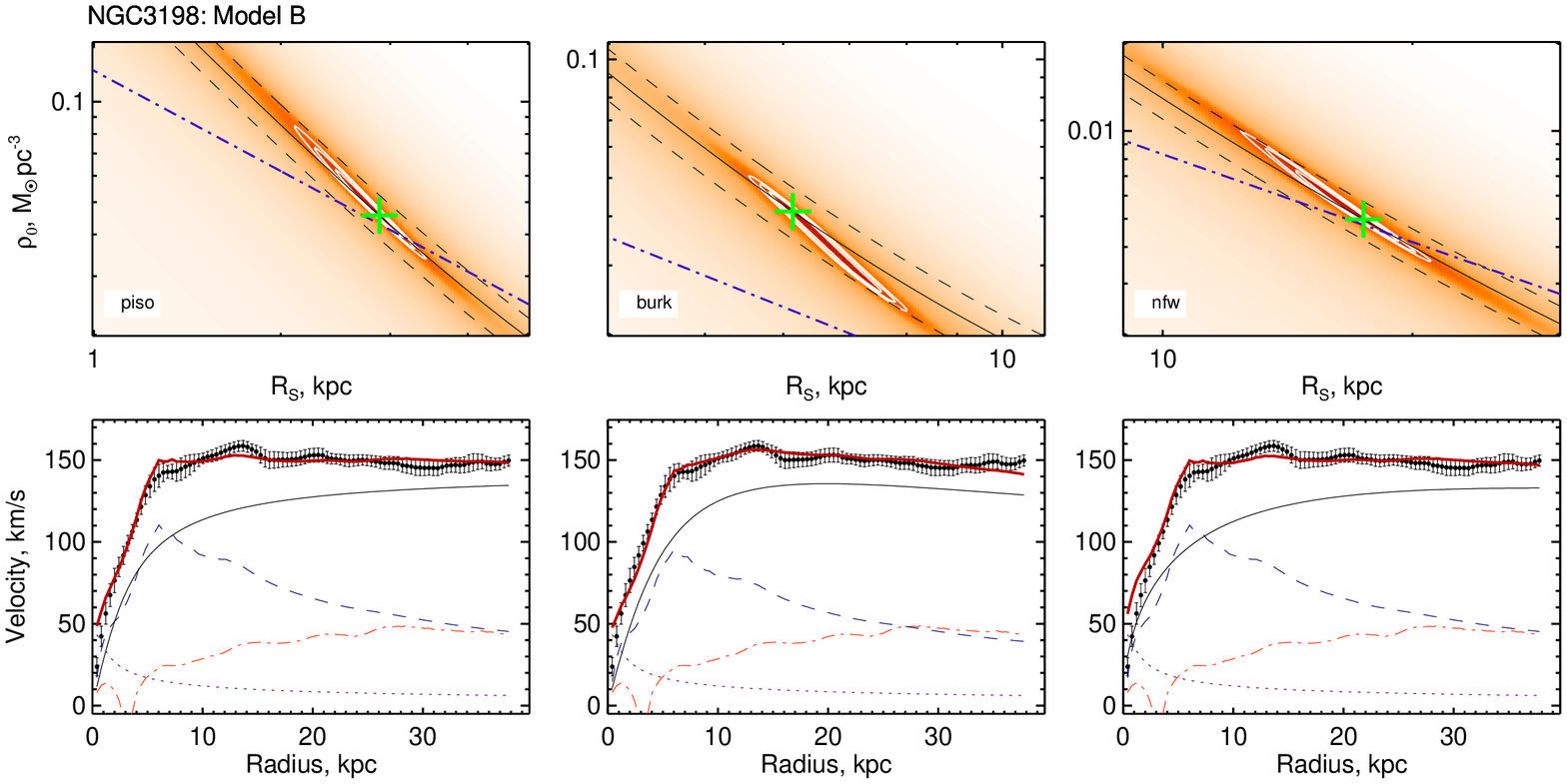}
\caption{(Continued).}
\end{figure*}

\begin{figure*} 
\includegraphics[width=16.5cm]{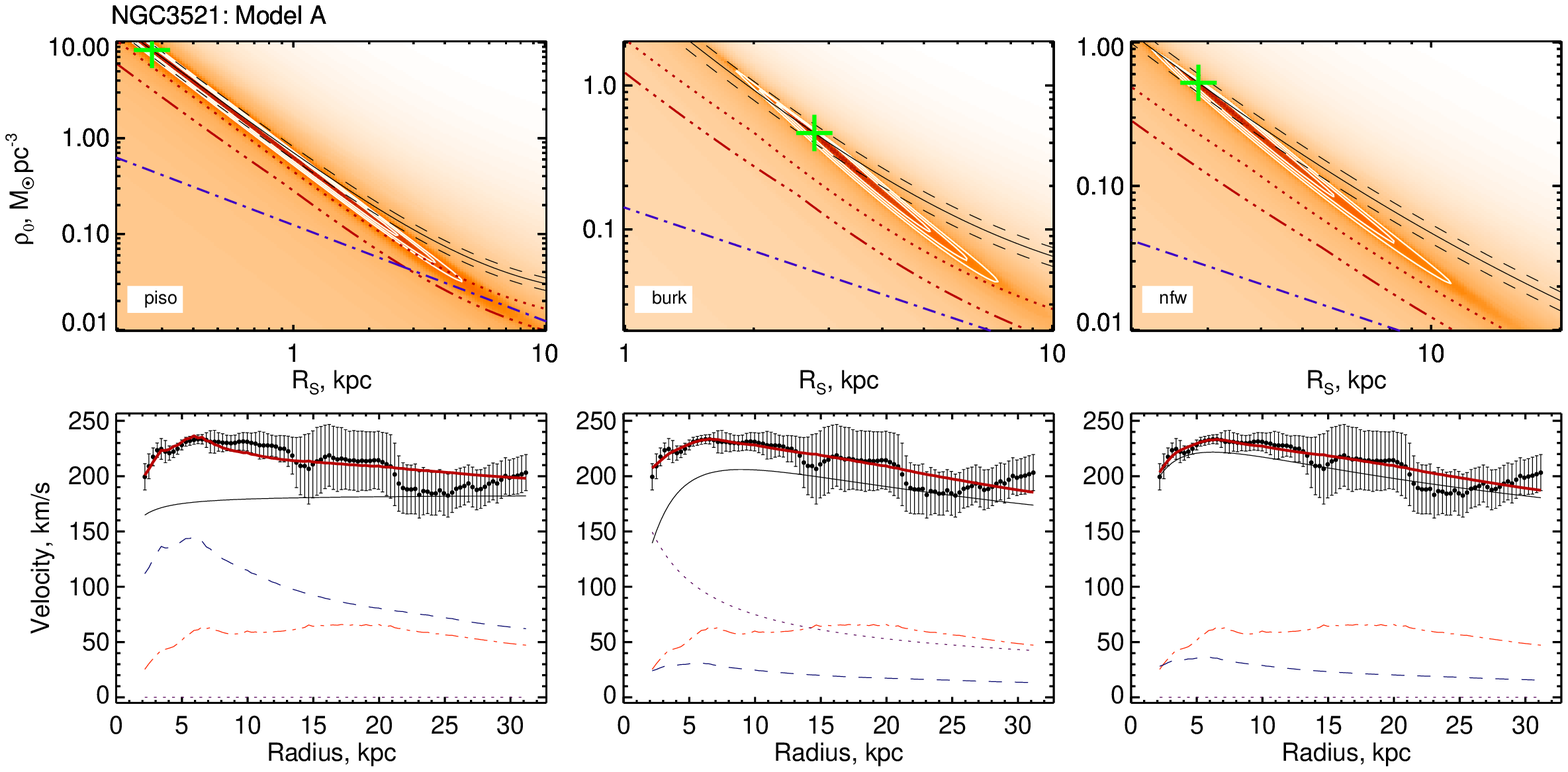}
\includegraphics[width=16.5cm]{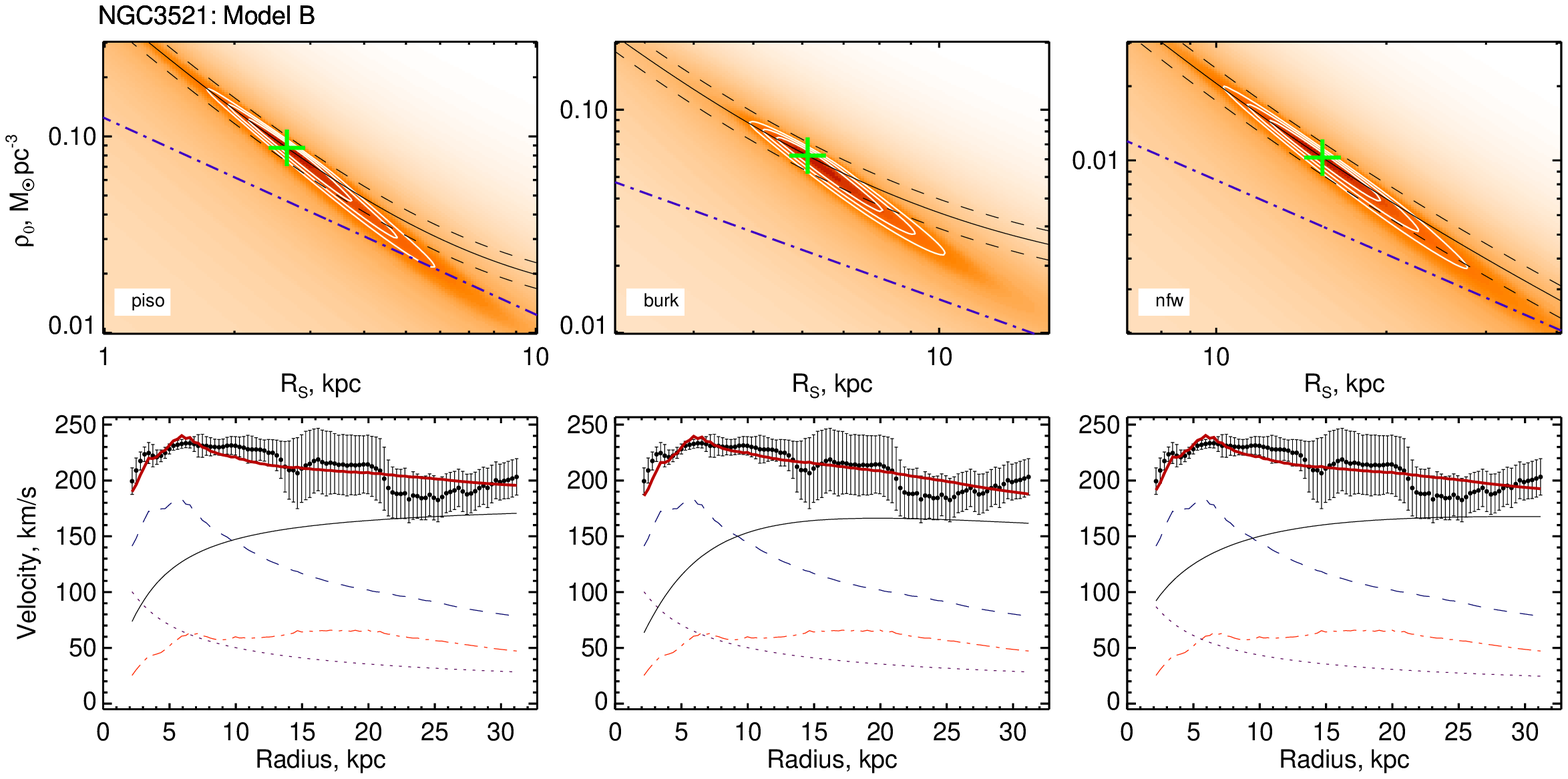}
\caption{The same as in Fig~\ref{fig1}, but for the cases, where only the plateau is present in the rotation curve.}
\label{fig3}
\end{figure*}

\addtocounter{figure}{-1}
\begin{figure*} 
\includegraphics[width=16.5cm]{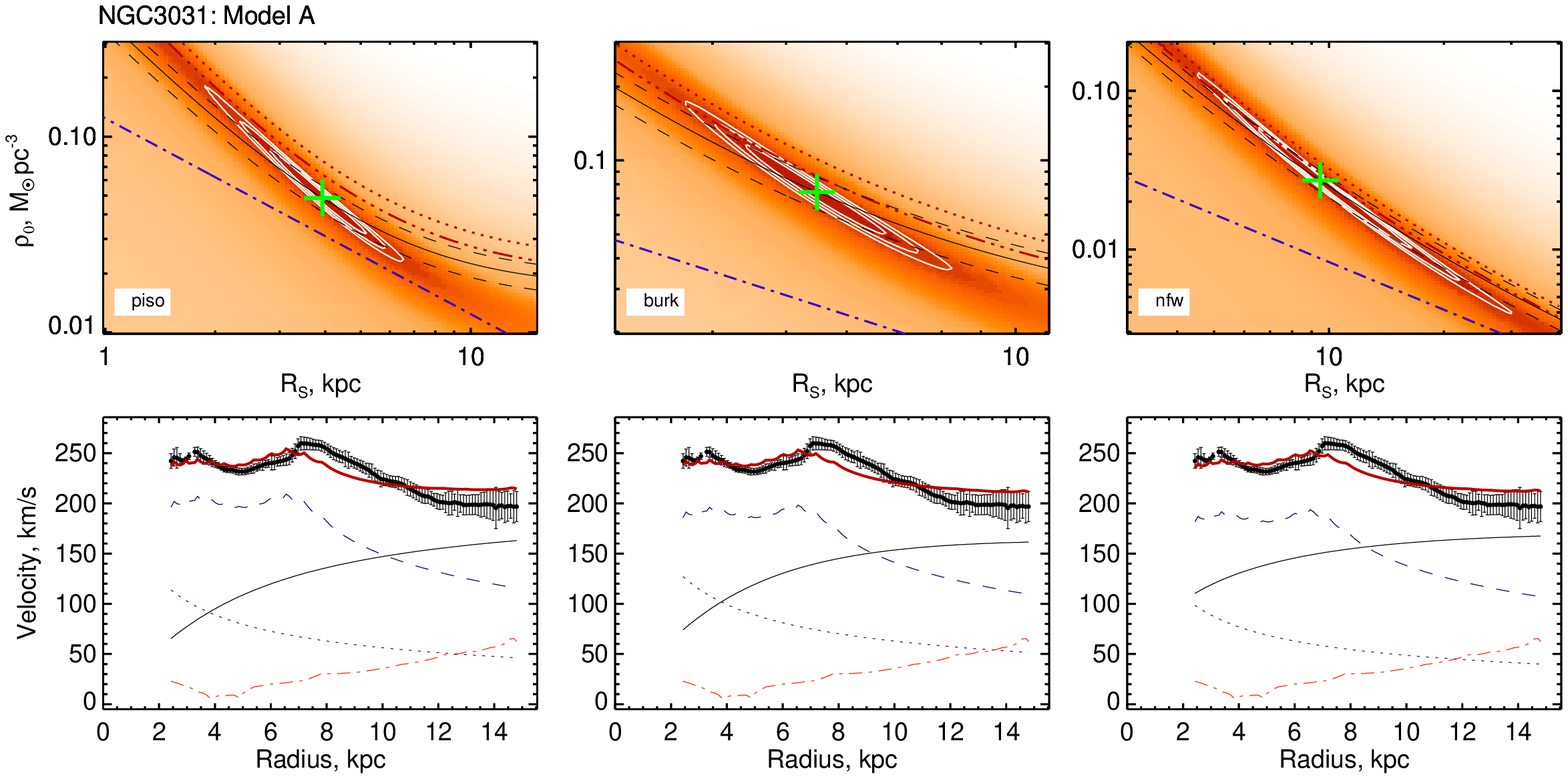}
\includegraphics[width=16.5cm]{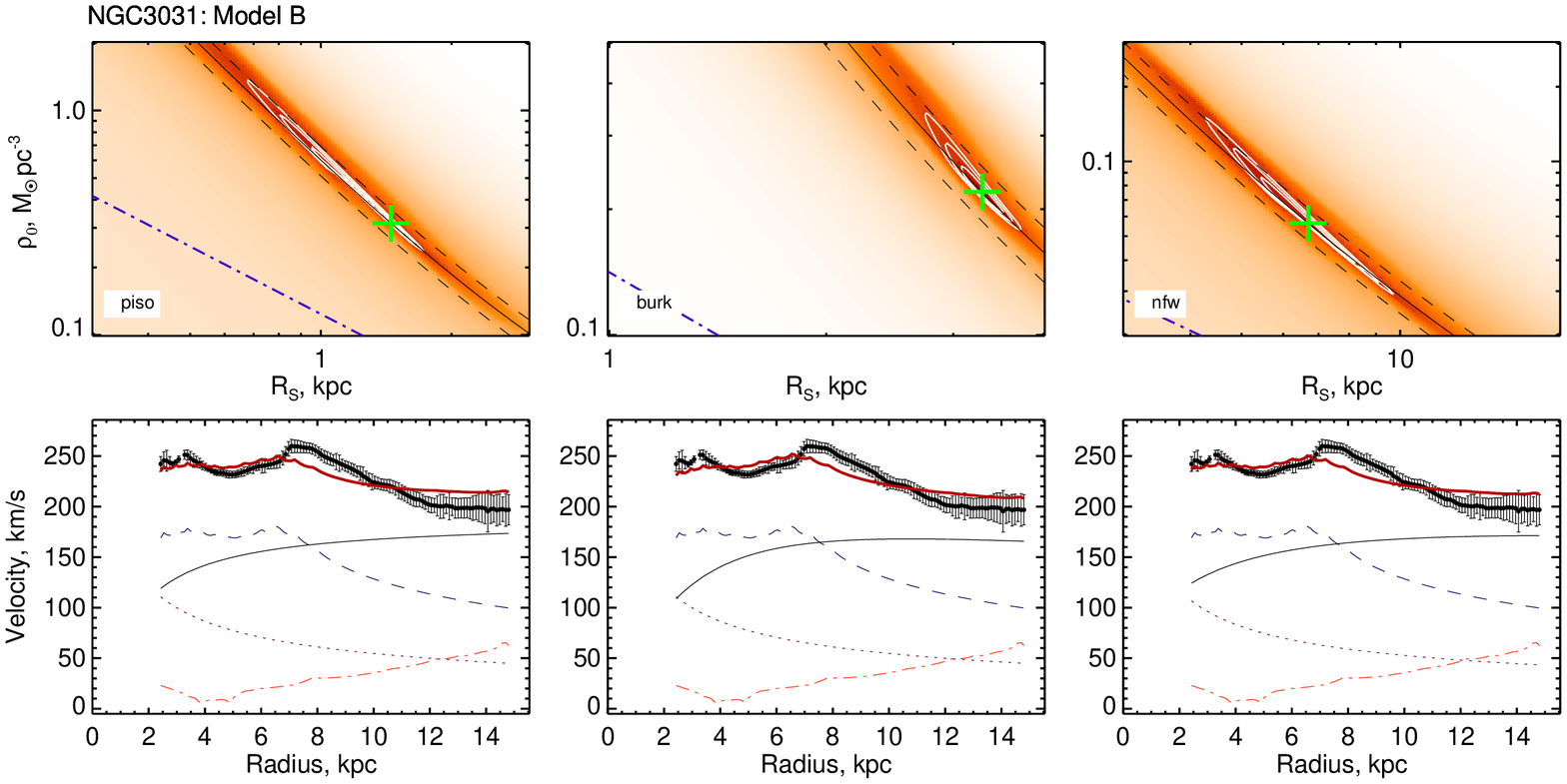}
\caption{(Continued).}
\end{figure*}

\begin{figure*} 
\includegraphics[width=16.5cm]{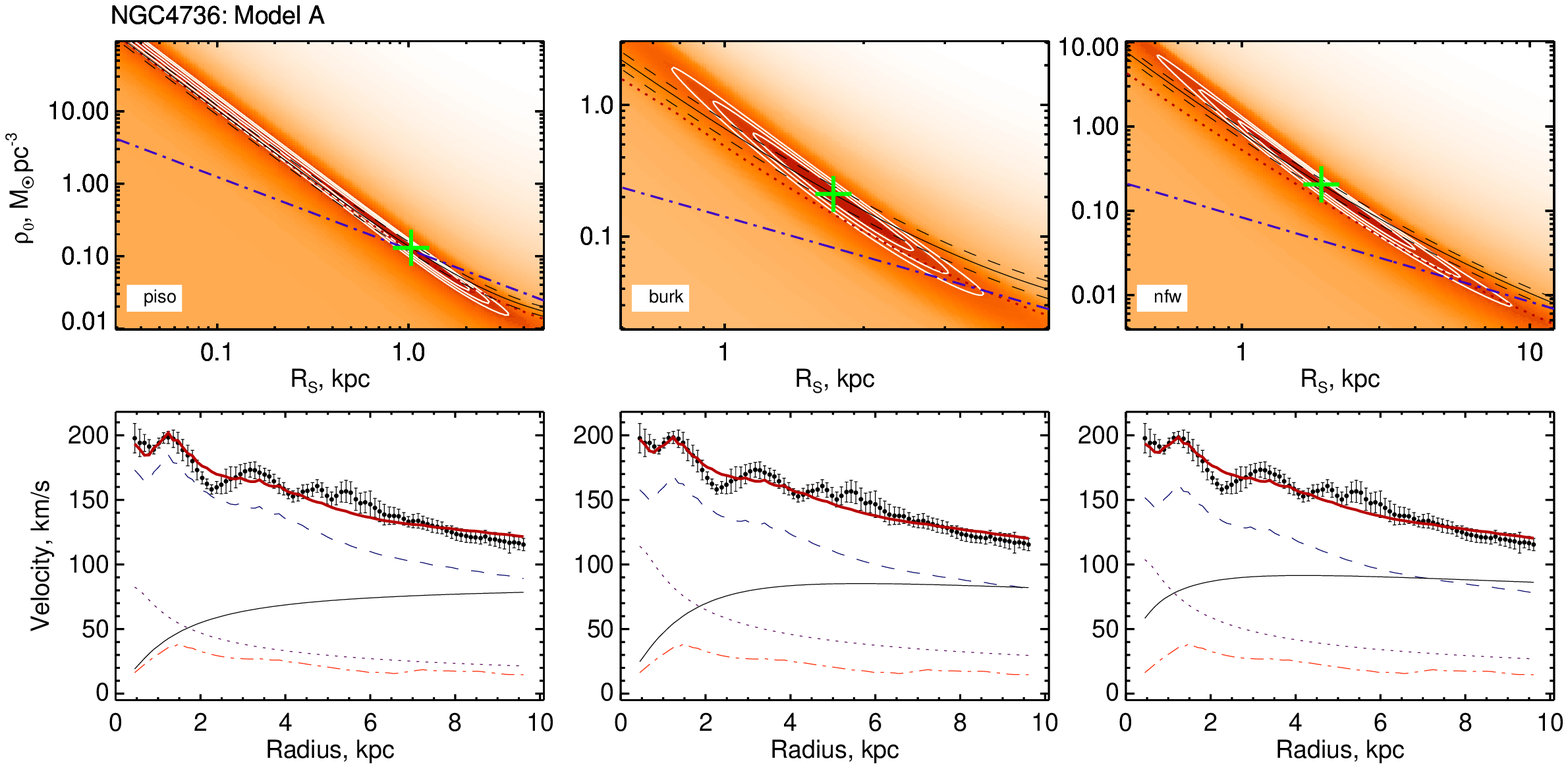}
\includegraphics[width=16.5cm]{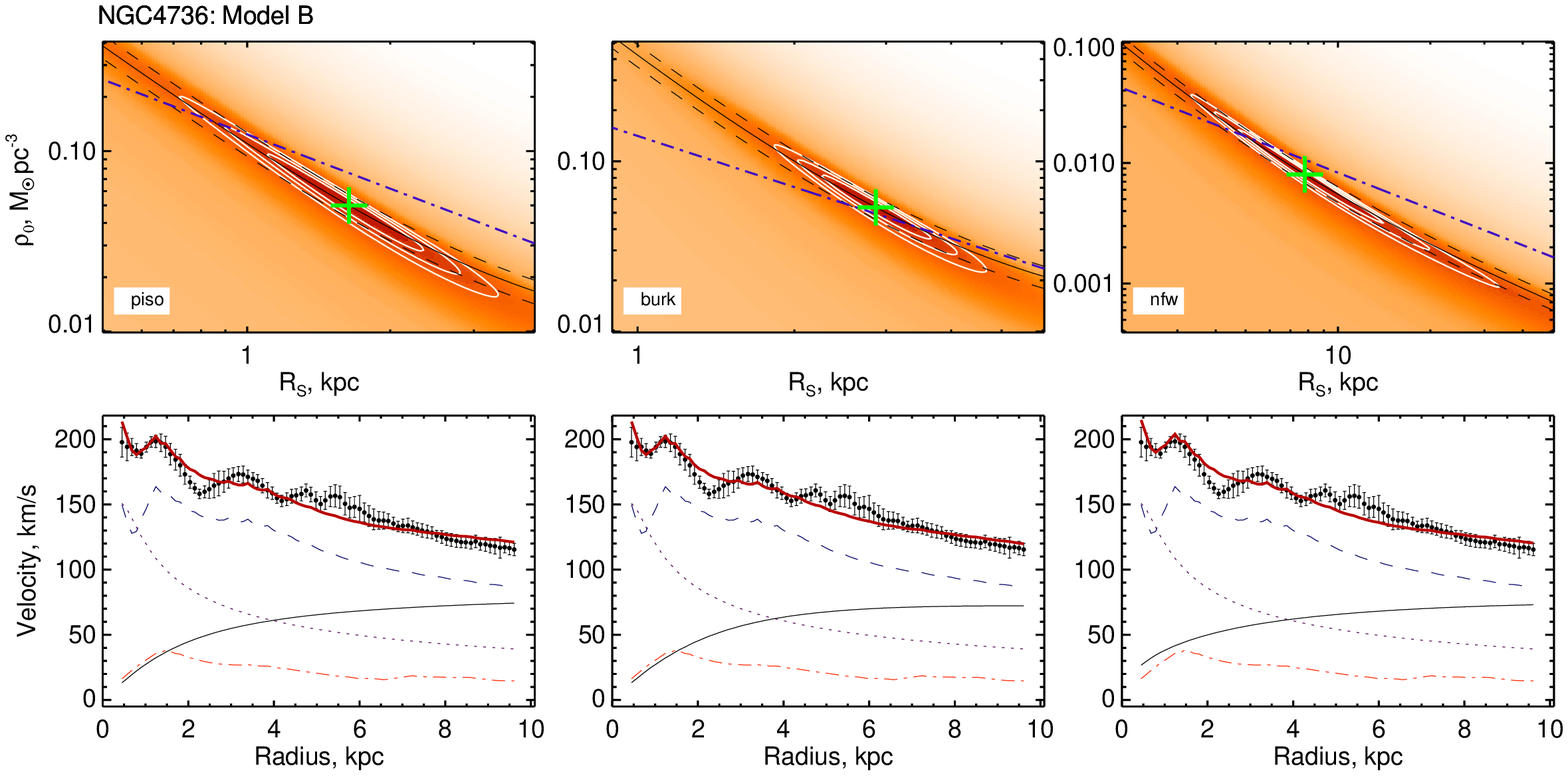}

\caption{The same as in Fig~\ref{fig1}, but for the case of falling rotation curve.}
\label{fig4}
\end{figure*}

\begin{figure*} 
\includegraphics[width=16.5cm]{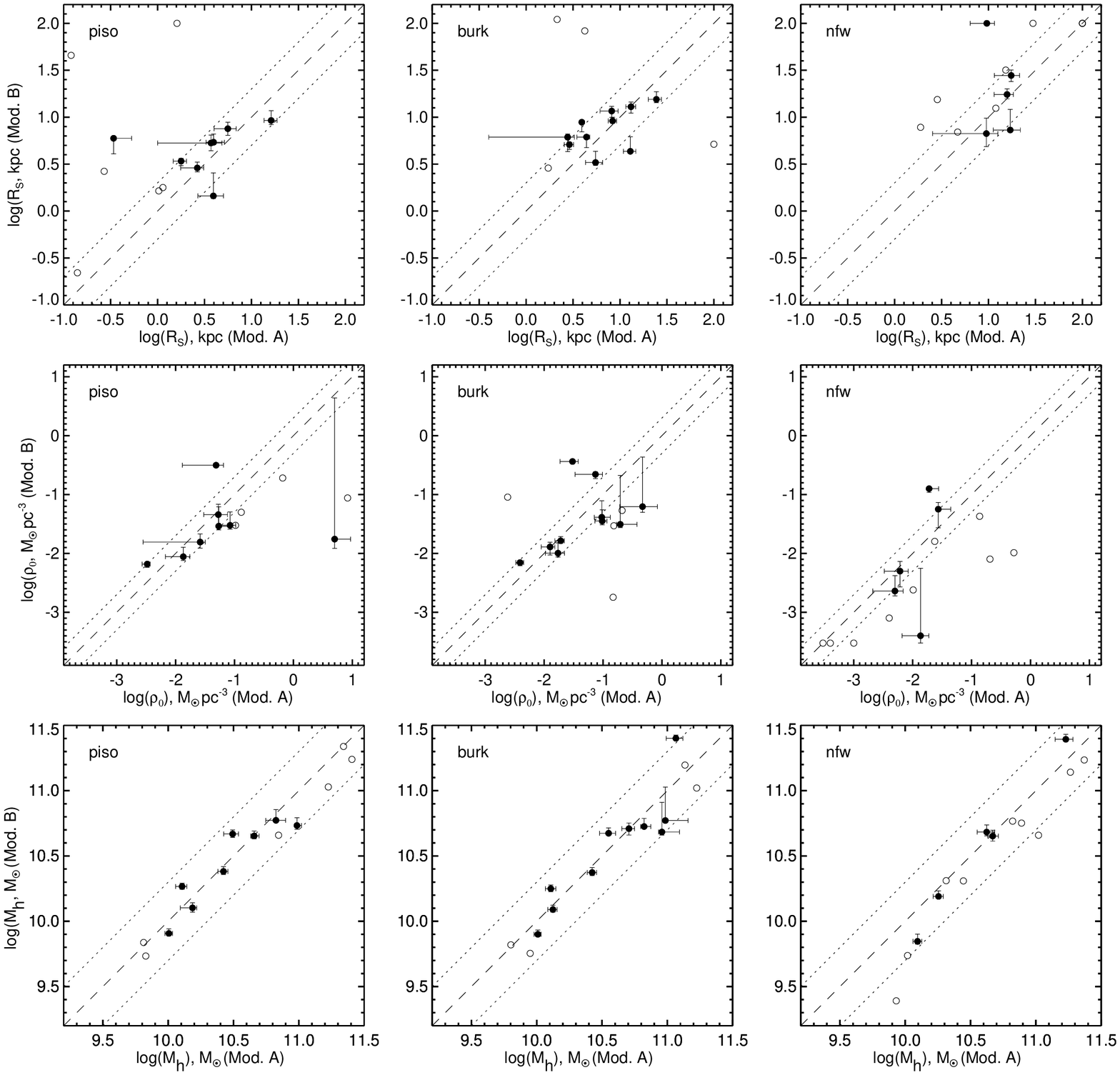}
\caption{Comparison of the results of photometrical and best-fitting modelling. The dashed and dotted lines correspond to the equal parameters and their twofold change. \green{Open circles mark the galaxies having one of the two models (A or B) for given halo profile with the infinite $1\sigma$ contours.}} 
\label{model_comparison}
\end{figure*}

\section{Results}
\label{results}

\begin{table}
\begin{center}
\caption{The evaluation of the reliability of the results of best-fitting \green{Models~A} for three halo profiles for different types of rotation curves. \label{kinds}}
\begin{tabular}{lccc}
\hline
NGC & piso & burk & nfw \\
\hline
\hline
\multicolumn{4}{l}{i. Increasing rotation curve}\\
\hline
0925	&	+ & + & $-$ \\
2976    & + & + & $-$   \\
\hline
\multicolumn{4}{p{4cm}}{ii. Detailed data is present for inner parts~of rotation curve}\\
\hline
2403	& $+$  & + & +  \\
3198	& + & + & +  \\ 
3621	&	+ & + & $-$ 	 \\ 
5055	&	+	&	+ &  \green{+} \\ 
7793	&	$+/-$	&	$+/-$ & $+/-$	\\ 
\hline
\multicolumn{4}{l}{iii. Only plateau}\\ 
\hline
2841	& $-$ & $+$  & $+$ \\
2903	& $-$  & $+/-$  &$-$	\\ 
3031	& + & + & + \\ 
3521	& $-$  & \green{$+$} & \green{$+$}\\ 
6946	&	+	& $+$  & +\\ 
7331	&	$-$ 	&	$-$  & $-$ \\ 
\hline
\multicolumn{4}{l}{iv. Falling rotation curve}\\
\hline
4736	&	$-$  & \green{$+$}  & \green{$+$} 	\\ 
\hline
\end{tabular}
\end{center}
\end{table}

\subsection{\green{Comparison of the models}}\label{comp}

\green{The interconnection between the DM halo parameters represents a~separate important problem. 
As one can see from Figs.~\ref{fig1}--\ref{fig4} for both models the $\chi^2$ maps show similar degeneracy of the density and scale of dark halo, but in case the photometrical approach (Model~B) the spread of the parameters is significantly lower.
Despite of the lower estimation errors (see Tables~\ref{parameters_burk}--\ref{parameters_nfw}) of the dark halo parameters the photometric method has its own weak point. }
The reduced values of the $\chi^2$ are significantly higher than for the best-fitting technique, which is expectable, but makes these models less reliable. 
\green{Moreover, for the three galaxies (as mentioned in Sect.~\ref{ml}), we were unable to get a satisfactory fit for the standard range of M/L at all. 
In some cases the M/L range extension does not improve a situation} \green{(see Fig.~\ref{fig2} for \N7793). This problem was also mentioned in \cite{THINGS}\footnote{See Figs. 30, 37, 48, 50, 52, 53 for fixed baryonic surface density in the cited paper.}.}
\green{Thus despite of the restriction of mass-to-light ratio decreases the dispersion of the possible parameters of DM halo it does not make the results of the mass-decomposition more reliable in many cases. }

\green{Another problem is the existence of models with infinite contours of $1\sigma$ confidence levels. \green{In other words, in these cases, the rotation curve can be described equally well by using a halo with the cardinally different parameters.  }
\green{Even when the baryonic surface density is fixed we still got such models\footnote{\green{For \N0925 (NFW), \N2976, \N2841 (piso), \N2903 (piso, Burkert), \N3621 (NFW), \N6946 (NFW), \N7793 (NFW).}}.
Noticeably models with NFW profile fail more often than that with other profiles when the baryonic surface density is fixed.  }}  

\green{In Fig.~\ref{model_comparison} we demonstrate the comparison of central densities, radial scales and DM masses within optical radius  resulting from \green{Models~A and B}. 
\green{Open circles mark the galaxies for which one of the two models for a considered halo profile give the infinite $1\sigma$ contours. 
The dashed and dotted lines correspond to the equal parameters and their twofold change.}
As one can see,  \green{the results of DM halo mass determinations are in good agreement for Models~A and B}. 
Moreover, even the \textit{unreliable} models (with infinite $1\sigma$ contours) give the stable mass estimations, due to the degeneracy between $R_s$ and $\rho_0$ along the line of constant halo mass.
} \green{The solutions for the central density and the scale are much more sensitive to the choice of the baryonic surface density.}

\subsection{Comparison with other authors}

\green{We compared the results of our best-fitting mass-modelling with the dark halo parameters found in \citet{THINGS}\footnote{We took their estimations for the best-fitting model, if these results were missing we used photometrical approach.}, \citet{CardoneDelPopolo2012}, \citet{Kasparova2012}, \citet{Frank2015} and \citet{Karukes2015}. We found a satisfactory agreement for most of the cases. But for some galaxies the discrepancies exceeded the 3$\sigma$ confidence limits (\N5055, \N7793, \N3621, \N0925 for Burkert halo). In most cases the discrepancies are most likely related to the difference between the photometrical data from the present and other papers. }

Another cause of the discrepancy of the estimations is that the observed rotation curve is badly reproduced by given type of halo density profile and it could lead to big uncertainties of the DM halo parameters. 
For example, for \N0925 the Burkert halo density profile gives the bad fit to the rotation curve according to \citet{CardoneDelPopolo2012}.

\subsection{Halo masses}

\green{Our results show that in most of the cases the $\chi^2$ minimum regions lie along the line of the constant dark halo mass and the spread of the dark halo parameters corresponds to the range of $M_{halo}\pm15$~per~cent. 
Thus, for the best-fitting modelling the estimation of $M_{halo}$ is much more accurate than that of $\rho_0$ and $R_s$.
As it was discussed above, in case of the baryonic density fixation it can just \green{follow}\ from a halo shape. 
However, this relation becomes less evident for the best-fitting modelling without any limits on the mass-to-light ratios of disc and bulge when the stellar density can vary by several times.} 

\red{The constancy of dark halo mass could possibly arise from the narrow range of the dark halo contribution to the rotation curve $v_{halo}(r)$ at the outermost regions.
If the halo velocity at optical radius is close to the observed velocity at the outskirts of a galaxy $v_{halo}^2(R_{opt})\approx v^2(R_{max})$, we can get approximate estimate of the dark halo mass which is almost identical to the dynamical mass inside of $R_{opt}$: $M_{dyn}= v^2(R_{max})R_{opt}/G\approx v_{halo}^2(R_{opt})R_{opt}/G$. In order to test this possibility, we calculated this rough estimate of dark halo mass and compared it to the values followed from the rotation curve decomposition. We give the ratios of the estimates in column 9 of Tables ~\ref{parameters_burk}--\ref{parameters_nfw}. 
As one can see from the Tables despite there are some cases when the rough estimate is close to that found from the rotation curve decomposition there is a significant number of galaxies for which the ratios of the dynamical masses and the halo masses found from the rotation curve decomposition differ significantly from unity.
For roughly a quarter of the sample the ratios of the masses exceed 2. 
In a majority of cases the dynamical estimate is higher than follows from the mass-modelling which is expectable due to the neglecting of the baryonic matter contribution to the rotation curve at the outer radius which appears to be different for different galaxies. 
However, in a few cases the ratio of masses is lower than one, which indicates that the contribution of the dark halo at the optical radius in the model is higher than the velocity at the outermost point of the rotation curve. 
It is possible when the rotation curve has a bump followed by the lower rotational velocity amplitude in the plateau. 
In this case the mass-model traces the shape of the rotation curve which is impossible for the rough estimate. 
Another possibility is when the rotation curve does not reach the plateau making the estimate of $v_{halo}^2(R_{opt})$ uncertain.}

\red{Having in mind all described problems of the rough estimate of the dark halo mass we conclude that the rotation curve mass-modelling is still needed to obtain the reliable mass of the dark halo.  }

\green{It should be noted that in the case of Models~A}, there are some exceptions in which the $\chi^2$ minimum behaves differently in comparison to the line of equal mass. 
In 6 out of 42 models~--- \N2903 (Burkert), \N3521, \N6946 (Burkert), \N7331 (piso)~--- $1\sigma$ confidence limits form two separate isolated contours, corresponding to two different values of DM halo mass. In these cases the two different total halo masses are possible, since there are two separate minima on the $\chi^2$ which lie in the two spaced and parallel lines of constant DM halo mass.

\green{In spite of this for the majority of instances} we can make the following general conclusion. \green{If the study is concerned on the dark halo masses, and is not aimed on its density, radial scale or \green{concentration}, the pure best-fitting mass-modelling of the rotation curve could be enough for this purpose.}

\green{In addition to the above it is worth mentioning that the usage of NFW profile gives the halo mass estimations systematically higher (up to $\sim15$~per~cent) in comparison with other two halo shapes} for most of the galaxies of our sample.

\subsection{The disc mass-to-light ratio}

The disc mass-to-light ratio can give important information that narrows the range of the dark halo parameters estimated from the mass-modelling. 
From Tables \ref{parameters_burk}---\ref{parameters_nfw} it is evident that \green{in the best-fitting approach} in contrast to the $M_{halo}$ the mass-to-light ratio of the disc is much more uncertain (the error exceeds 100~per~cents in some cases). 

We plotted on all $\chi^2$ maps (for Model~A)  the red lines of constant disc mass-to-light ratios\footnote{\green{In some cases (\N0925, \N2403 for Burkert halo) the pure best-fitting model gives the mass-to-light ratios that differ significantly from the photometrical ones, thus the red lines lie beyond the limits of the maps. }} \green{coinciding with the limit values for the Model~B given in Sect.~\ref{ml}.}
As we can see from Figs.~\ref{fig1}--\ref{fig4}, in some cases one value of $M/L$ corresponds to the set of pairs $\{\rho_0, R_s\}$ (see e.g. Model~A in Fig. \ref{fig2} for \N3198). 
Another problem is that in some galaxies from the sample the result of \green{Models~A} is associated with $M/L$ that are too low or too high in comparison to the value expected from the photometry. 
It could indicate that either the part of mass is incorrectly assigned to some of the galaxy components or that the disc is more/less massive than it follows from the photometry. 
The later reason seems more questionable.

\begin{figure*} 
\includegraphics[scale=0.73]{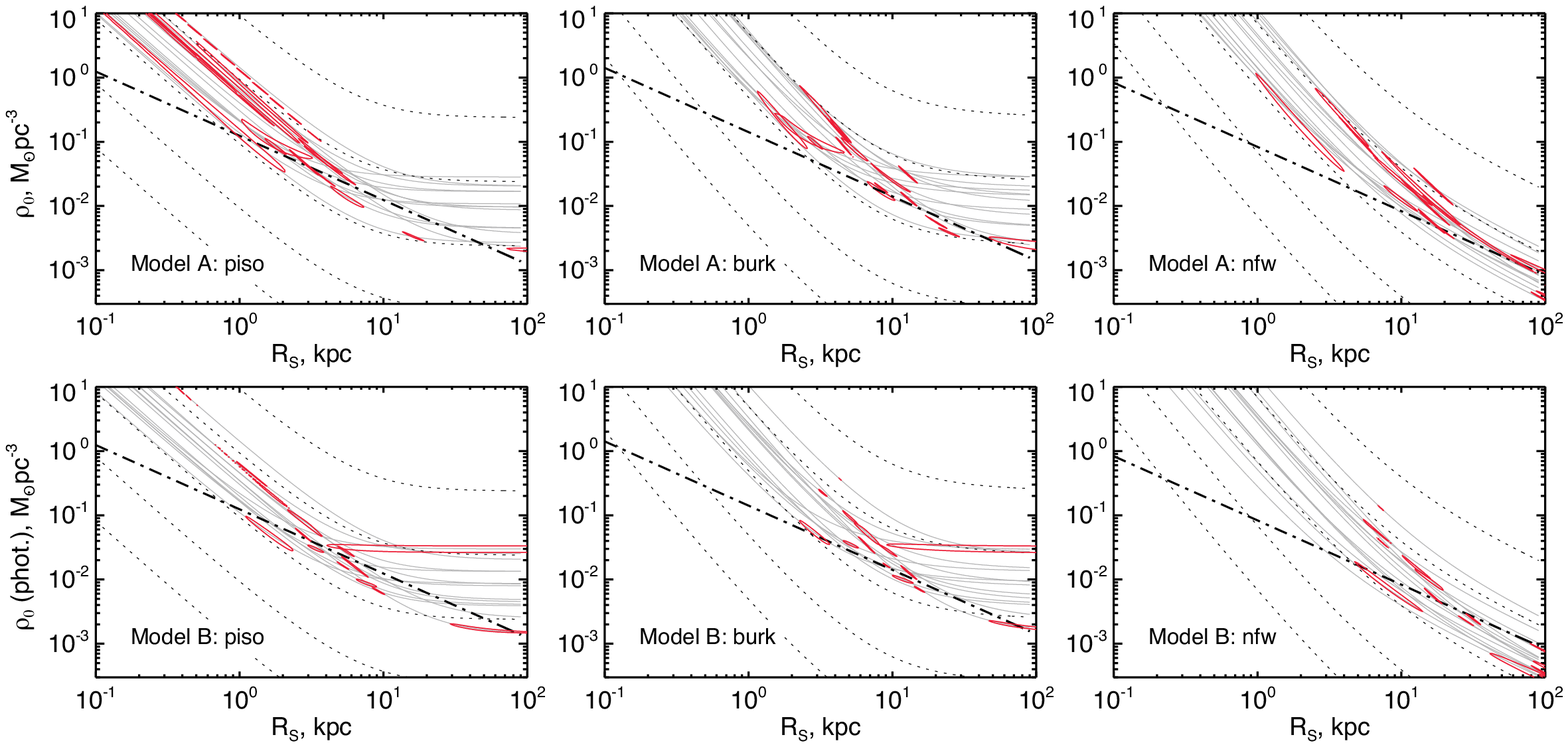}
\caption{
\green{The $1\sigma$ confidence limits of all galaxies for three types of dark haloes for Models~A (top panels) and Models~B (bottom panels).   
The dash-dotted lines denote $\log(\mu_{0D}) = 2.15$. 
Dotted curves mark the lines of constant mass within $R_{opt}=10$~kpc: $\log (M_{halo})=8$, $9$, $10$,  $11$ and $12$ (from bottom-left to top-right of each panel) in solar units. 
Gray lines correspond to the constant mass lines of the galaxies of our sample.}}
\label{dm_parameters}
\end{figure*}

\subsection{\green{Analysis of Model~A results based on the characteristics of~rotation curves}}\label{classification} 

One of our aims was to understand how the features of a rotation curve influence the results of the mass-modelling. We decided to do it for the most general model (Model~A).  

\green{We found} from the $\chi^2$ maps (together with the parameters limits given in Tables~\ref{parameters_burk}--\ref{parameters_nfw}) that not in every case the results of Models~A are reliable. 
We developed the following criteria in order to unveil the \textit{untrustworthy} models: 
(a)~the lack of separate minimum on the $\chi^2$ map (the infinite $1\sigma$ contour); 
(b)~the solutions with zero disc contribution to the rotation curve. 

For the convenience we arbitrarily divided the galaxies into the subsamples with different types of input kinematic data.
Thus, taking into account the behaviour of the $\chi^2$ map and best-fitting models, we assert the importance of the four features of rotation curves.

\begin{itemize}
\item[i.] The increasing rotation curve. 
Only two galaxies of our sample (\N0925 and \N2976) do not have a  plateau on the rotation curves, their rotational amplitudes are lower than 120~km/s. 
The rotation curves of \N0925 and \N2976 are not extended enough (the outermost radius is $<R_{opt}$ in contrast with other galaxies of the sample which have more extending rotation curves).
In both cases, the \green{Models~A} for NFW halo ignore the disc presence (in the case of \N0925 the disc mass-to-light ratio also equals to zero inside the entire $1\sigma$ contour of $\chi^2$ map, for \N2976 there are solutions with $M/L \le 0.04$ which is quite low).
The other DM halo profiles work correctly for these galaxies.
Probably, this is due to a specific shape of the NFW profile 
and, at the same time, because of the lack of a certain rotation curve plateau.
The latter does not allow to lock the parameters of the halo in the finite $1\sigma$ contour on $\chi^2$ map. 
It is worth noting that as it was previously shown by \cite{Kuzio2008}, the contributions of NFW profile and maximal exponential disc into rotation curves are mutually exclusive for LSBs. We strengthen this conclusion for HSBs with increasing rotation curves.

\item[ii.] Rotation curves with detailed data in the inner parts (including region within $1-2$~kpc). For this type of rotation curves on average, all models give satisfactory results. 
\green{Exceptions are the unsuccessful models with all halo types which are insensitive to the velocity fall on the disc periphery of \N7793}. 
\green{Besides it, though} the best-fitting model seems reasonable at first sight, it is impossible to obtain the NFW halo parameters of \N3621 (infinite $1\sigma$ contour).

\item[iii.] The rotation curve with a lack of data in a few inner kiloparsecs (it could be probably due to the depression of  H{\sc i} surface brightness in the central regions of these galaxies). 
In this case \green{almost half of} models do not work well. 
The majority of best-fitting models give very different estimations of the dark halo masses and the disc-to-halo mass ratios for three different DM halo profiles.
\green{Burkert and NFW models of \N3521 give a zero disc solution at almost the entire $1\sigma$ area.}
\green{There are two} galaxies of this category (\N3031 and \N6946)\footnote{\green{We did not take into account in our analysis that \N3031 has strong streaming motions along the spiral
arms and \N6946 has a low inclination angle which leads to high uncertainty of the rotation curve. It can possibly influence the accuracy of the input data but will not change the main results of our analysis.}}, which have good estimations of the dark matter parameters \green{probably} due to the special features of their kinematic data.
The first galaxy has a prominent bump on the rotation curve (at $7-8$~kpc) correlated with the stellar disc profile.
The second one has a very small bulge \citep{THINGS}, which is reflected in the slow growth of rotation velocity within 5~kpc.

\item[iv.] Decreasing rotation curve. This type is represented only by \N4736. 
\green{Our modelling of this galaxy \green{is very sensitive} to the photometric data. 
Even small changes in the input data can lead to the \green{zero disc solution}. 
As a result, we suppose that the method \green{should be used with caution}  for galaxies with decreasing rotation curves.}

\end{itemize}

In Table~\ref{kinds} we show the classification of our models taking into account the types of the input data. 
For clarity we use the following denotation: 
we put ``$-$'' for models that we think are unreliable (see above), 
``$+$'' marks good models, 
we mark ``$+/-$'' sign, if, from our point of view, the best-fitting \green{Model~A} gives an incorrect contribution of one of the components due to not sufficient extension of the rotation curve (e.g., in the case of \N7793).

As one can see from Table \ref{kinds}, models with various profiles of DM halo are differently sensitive to the characteristic features of the input kinematic data.
Applying the Burkert halo gives good results for \green{11} of the 14 galaxies. 
\green{In cases of the piso and NFW profiles~--- 8/14 good and 5/14 false models for both.}

\section{Discussion}
\label{discussion}

\subsection{The dark halo surface density}

To test \green{the behaviour of} dark halo surface densities and to understand how uncertain they could be we \green{plotted the $1\sigma$ confidence limits of the $\chi^2$ maps for all our galaxies together.}
In Fig.~\ref{dm_parameters} we show the central densities versus the radial scales for three types of dark haloes \green{for Models~A (top panels) and Models~B (bottom panels)}.  
The dash-dotted lines correspond to $\log(\mu_{0D}) = 2.15$ found by \citet{Donato2009}, corrected for given type of dark halo density profile as shown above (Eq.~\ref{eq9}--\ref{eq10}). 
\green{Dotted curves mark the constant mass lines for examples of objects with $R_{opt}=10$~kpc and $\log (M_{halo})=8$, $9$, $10$, $11$ and $12$ (from bottom-left to top-right of each panel) in solar units. 
Gray lines correspond to the constant mass lines of galaxies of our sample.} 

\green{We can see that it is impossible to make valid conclusion about the universality of dark halo surface density using mass-modelling of the rotation curve in spite of the high quality of input data due to the big uncertainty even for Models~B. \green{The possible exception can be made for the diagrams for models with Burkert profile which are more stable than the others, so the estimates are less uncertain. Despite of the wide ranges of the DM halo parameters and a small number of the galaxies we can see that there are several galaxies that have DM surface densities higher than was found by \citet{Donato2009} and the overall scatter of points is high.}}

\green{ Another conclusion, that we draw from Fig. ~\ref{dm_parameters} is that as far as we consider here the sample of galaxies with the narrow range of halo masses we observe the steeper artificial correlation related to } the degeneracy between $\rho_0$ and $R_s$ along the lines of halo mass constancy. Similar artificial correlation for LSBs was found and discussed e.g. by \cite{deBloketal2001}.

\begin{figure*} 
\includegraphics[scale=0.65]{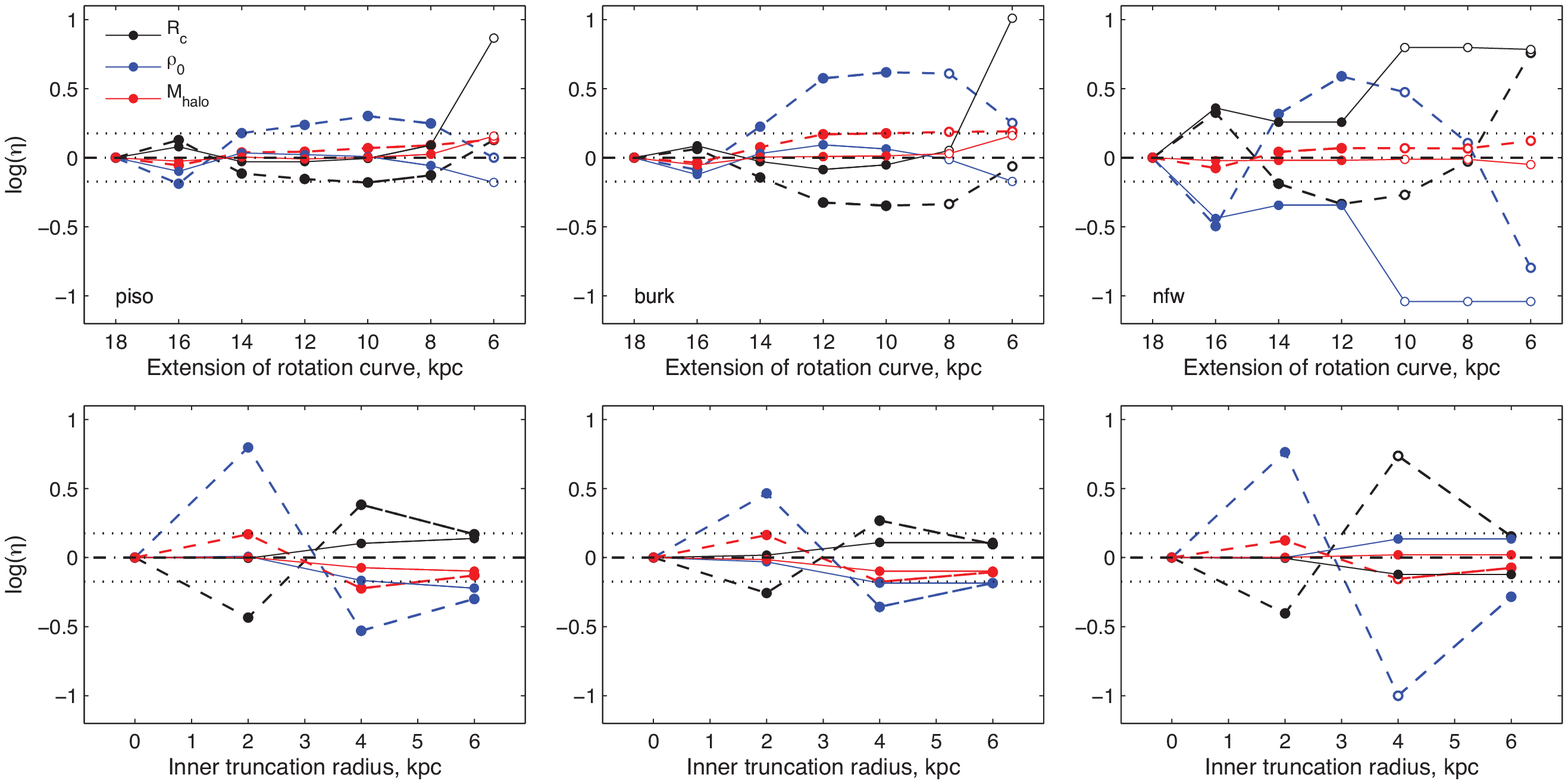}
\caption{The results of test of the effect of rotation curve truncation for \N2403. 
The top panels show the relative changes $\eta$ of parameters $R_s$, $\rho_0$ and $M_{halo}$ (normalized to the ones for the models of non-truncated observed rotation curve) 
as a function of the maximum radius with \textit{known} rotation velocity.
Dashed lines show the best-fitting Models~A, solid lines~--- Models~B.  
Open symbols mark the untrustworthy models with infinite $1\sigma$ contour.
The dotted lines limit the change of parameters in the range of $\pm50$~per~cent.
In the bottom panels we show the same but for the case of the inner part truncation of rotation curve from $r=0$ to 6~kpc.}
\label{fig1_cut}
\end{figure*}

\subsection{On the effect of data quality on the mass-decomposition}
\subsubsection{The influence of the rotation curve extension on~the~modelling}

To test the sensitivity \green{of both Models A and B} to the quality of the input kinematic data, we made two types of \textit{experiments}.
In the first type we chose several galaxies with sufficiently extended rotation curves and good mass-models.
Then we cut the part of their observed rotation curve, as if these data are not available.
After that, we run our program of the rotation curve decomposition using the truncated data and look at the changes of the parameter estimations of all halo types (piso, Burkert and NFW) \green{for both pure best-fitting Models~A and photometrical Models~B}.  We did this test both for \green{the truncation of} inner and outer radii, \green{in order to simulate observational features} connected to the evolution of galaxies (the~H{\sc i} stripping from the disc periphery and the depletion of cold gas in the inner kiloparsecs).

We show in Fig.~\ref{fig1_cut} the results of our test for \N2403 as an example. 
The top panels demonstrate the logarithm of the relative changes $\eta$ of the parameters $R_s$, $\rho_0$ and $M_{halo}$ (normalized to the ones for the models of non-truncated rotation curve) 
as a function of the outermost radius of \textit{known} rotation curve (from 18 to 6~kpc).
Dashed lines show the best-fitting Models~A, solid lines~--- Models~B.  
Open symbols mark the cases of infinite $1\sigma$ contour and accordingly untrustworthy model parameters.
The dotted lines limit the change of the parameters in the range of $\pm50$~\green{per~cent}.
In the bottom panels we show the same plots but for the case of the inner part truncation of the rotation curve from $r=0$ to 6~kpc.

The diagrams for various galaxies look a bit differently, but we can make the following general conclusions:
\begin{itemize}
\item There is an obvious degeneracy between parameters $R_s$ and $\rho_0$ for all types of halo profiles. 
This conclusion also follows from the $\chi^2$ maps~--- the contours of $1\sigma$ confidence limits are elongated and inclined in the $\rho_0$ vs $R_s$ diagrams.
\item The $R_s$ and $\rho_0$ parameters can vary dramatically (by several times), even for a small changes of the rotation curve extension.
\item \green{
The truncation of the rotation curve at certain radius lead in some cases to the infinite  $1\sigma$ contour, which makes impossible the correct determination of $R_s$ and $\rho_0$.
It is seen for the models with NFW profile of \N2403 starting from the outer cutting at 10 kpc (this is about the middle of the rotation curve plateau of \N2403).}
\item \green{The changes of the parameters become more moderate (especially for the inner truncation) for Model B in comparison to Model A. In certain points, however, the changes for Model B are even higher than for Model~A (see e.g. upper row of Fig.~\ref{fig1_cut} for NFW profile). The influence of the data extension can be significant even for the models with fixed baryonic surface density. }
\item  The DM halo mass is the most reliable parameter that comes from the mass-modelling, this conclusion is strengthened by the analysis of $\chi^2$ maps. 
\end{itemize}

\green{In the best-fitting Models~A the} solution for the disc mass-to-light ratio $M/L$ and central surface density of bulge $(I_0)_b$ is also unstable. 
Namely, in the absence of the additional information on the densities of disc and bulge one can get  crucially different contributions of disc and bulge using the rotation curves of the same object but obtained, for instance, with different observational instruments. 
It can make the estimate of the parameters of dark halo even more unreliable.

 \begin{figure*} 
 \begin{center}
\includegraphics[scale=0.53]{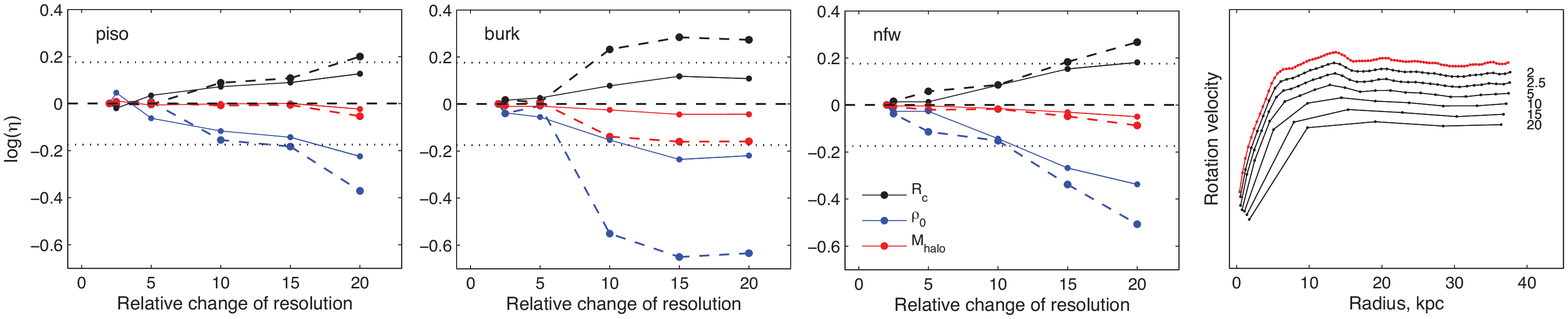}
\caption{The effect of resolution degradation for \N3198 for three types of DM halo \green{for Model A (dashed lines) and B (solid lines)}. 
The right panel shows a shape of the rotation curve for different resolution. 
The curves are artificially shifted for illustrative purposes. 
Initial curve is demonstrated by red line. 
The ratio of the new to initial resolution is marked near each curve.
On the right panels we show the relative changes $\eta$ of parameters $R_s$, $\rho_0$ and $M_{halo}$ (normalized to the ones for the models of the observed rotation curve) as a function of the ratio between the bin size and initial resolution of the rotation curve.
The dotted lines limit the change of parameters in the range of $\pm50$~per~cent.}
\label{resolution}
 \end{center}
\end{figure*}

\subsubsection{The effect of spatial resolution on the results of~mass-modelling}

In the second type of experiments we decided to test the influence of the spatial resolution on the results of rotation curve decomposition. This effect could be very important especially for distant galaxies. And it could be essential to know whether the estimates of densities and radial scales of DM haloes for distant galaxies can deviate from that of the nearby galaxies due to some systematical errors related to the effects of finite spatial resolution.

To make this test we chose again the galaxies with reliable mass-models. In order to get worth resolution, we convolved the rotation curves with Gaussian with FWHM equal to the ratio of the new and initial spatial resolutions. After that we changed the number of bins proportionally to the relative change of resolution. We got gradually larger and larger bins and ran our program for each resolution. 
In Figs~\ref{resolution} we show the effects of degradation of rotation curve resolution on the example of \N3198 for three types of DM haloes \green{in cases of the best-fitting Model~A (dashed lines) and the photometrical Model~B (solid lines)}. 
The \green{right} panel demonstrates the rotation curves for different resolution (the ratio of the resolutions is marked near each curve). 
Red line gives the initial rotation curve. In left panels we plot the logarithm of relative changes $\eta$ of parameters $R_s$, $\rho_0$ and  $M_{halo}$ (normalized to the ones for the models of the observed rotation curve) as a function of the ratio between the new and initial resolution of the rotation curve.
From Figs~\ref{resolution} one can see that the degradation of resolution has significant impact on the DM parameters, \green{which does not disappear if the baryonic density is fixed}. 
\green{The change of resolution has the highest effect on the estimates of $R_s$, $\rho_0$ (and hence on the dark halo concentration), that is consistent with the conclusions by \cite{deBloketal2001} on the influence of the data resolution on the concentration parameter of dark halo for models with the NFW profile.} 
In principle it can lead to the wrong conclusions if one compares the DM halo parameters of the nearby and distant galaxies obtained using the data with the same angular resolution. 
At the same time as in previous test the DM halo mass remains the closest to its value for the initial resolution.

\section{Summary} \label{Conclusions}

We showed that {in the case of mass-modelling of rotation curves} there are considerable \green{estimate uncertainties of the central densities $\rho_0$ and the scales $R_s$ (and thus the concentrations) of the dark halo, unlike its mass,} even using quite good input data.

We constructed the $\chi^2$ maps of the parameters of dark haloes for a sample of galaxies with high quality rotation curves taken from \citet{THINGS}. 
We considered three types of dark halo \green{ radial density profiles}: pseudoisothermal \green{(piso)}, Burkert and Navarro-Frenk-White \green{(NFW) }. \green{In our analysis we deal with two possibilities. In the first \green{case} the mass-to-light ratios of disc and bulge were taken as free parameters \green{(pure best-fitting, Models~A)}. In the second possibility they were confined by the narrow range followed from the models of stellar population \green{(photometric approach, Models~B)}.}

\green{The Model~A is worth consideration because it represents the most general case of rotation curve mass-modelling, it is suitable for a large number of objects. 
The halo parameters obtained in this way are still widely used, e.g. for comparison with the predictions of the evolution simulations. The Model~B has its disadvantages in comparison to Model A: it needs high quality photometrical data and uses assumptions on the stellar population properties such as stellar initial mass function.}

The analysis of the $\chi^2$ maps allowed us to make conclusions on the reliability of the estimates of dark halo masses within optical radii, their central densities, radial scales \green{(and hence concentrations)}. 
The main results are given below.
\begin{enumerate}

\item 
We showed that \green{in \green{one-third} of cases, the Models~A are not enough \textit{trustworthy} (see Sect.~\ref{classification}).   
The ability to determine reliably the halo shape depends critically on the features of the rotation curve (type and extension of rotation curve, the presence of detailed data for its centre parts).}

\item 
\green{If there are accurate enough kinematic data with detailed internal parts (including within $1-2$~kpc) the pure \green{best-fitting} Models~A give satisfactory results in most cases.}
In the case of the falling rotation curve or the poor data of inner parts, almost  all models give predominantly unreliable results in the absence of some characteristic features in the kinematic data (as for \N3031).
\green{The model with NFW profile gives unreliable estimates of $R_s$ and $\rho_0$ for the increasing rotation curve.}

\item 
\green{Models~A with the various halo forms are differently sensitive to the input kinematic data. 
Applying the Burkert halo gives good results for 11 of 14 galaxies. In cases of the piso and NFW profiles~--- 8/14 good and 5/14 false \textit{untrustworthy} models for both.}

\item  \green{In the photometrical \green{Models~B} the uncertainty} \green{of the halo parameters is significantly lower, \green{than in Models~A}. However, even when the baryonic density is fixed there are still cases of infinite $1\sigma$ confidence limit contours on the $\chi^2$ maps. Moreover, the results of the photometrical modelling correspond to systematically higher values of $\chi^2$ (in several cases Model B gives unsatisfactorily fits to observed rotation curve). }

\item \green{
The comparison of Model~A and B results shows that the halo mass estimates are consistent for each halo shapes in both cases, in contrast to $R_s$ and $\rho_0$. 
Moreover, the \textit{untrustworthy} models give stable  mass estimations, due to the degeneracy between $R_s$ and $\rho_0$ along the line of constant halo mass.
Therefore, we can estimate successfully the halo mass using the pure best-fitting method without any restrictions on the mass-to-light ratios.
}

\item 
We tested the method sensitivity to the quality of the input kinematic data (extension and spatial resolution of the rotation curve). In our tests we modified the input data by analogy with the observation characteristics, that can make the data less detailed (the H{\sc i} stripping from the disc periphery, the depletion of cold gas in the inner kiloparsecs and poor spatial resolution \green{of data} for distant galaxies). 
Our experiments show that, \green{unlike the halo masses, the estimates of $R_s$ and $\rho_0$} can vary significantly, even in response to small changes of the rotation curve extension. 
The degradation of the spatial resolution has \green{noticeable influence on the resulting parameters too}. 
\green{All these effects persist even if the baryonic density is fixed, though become more modest.}
It could be essential if one tries to compare the halo \green{concentrations} for nearby and distant galaxies using the data with the same angular resolution. 

\item For most of our $\chi^2$ maps the contour of $1\sigma$ confidence limit lies along the line of constant halo mass and is covered by the uncertainty of the halo mass of 15~per~cent. 
\green{Therefore, taking into account the results of our experiments too,} the halo mass \green{within optical radius} is the most reliable parameter in the mass-modelling \green{even when the dark-to-luminous mass ratio is varied by several times in the inner region of a galaxy. }

\item  There is an obvious degeneracy between parameters $R_s$ and $\rho_0$ for all types of halo profiles, which is evident both  from the analysis of $\chi^2$ maps and from our tests of the method sensitivity \green{for Models~A and B}.
\green{So, we are convinced that it can be impossible to obtain the truthful estimates of the halo concentrations from observations using the mass-modelling of rotation curves.}

\item 

Our research has shown the impossibility to come to valid conclusion about the universality of dark halo surface density using mass-modelling in spite of the high quality of input data.
\green{Instead of it, there is a steeper artificial correlation $\rho_0(R_s)$   related to the degeneration of $R_s$ and $\rho_0$ along the lines of constant halo masses.}

\end{enumerate}

\section*{Acknowledgements} 
We thank the anonymous referee for the valuable critical comments. 
We are thankful to Anatoly Zasov and Ivan Zolotukhin for valuable comments on the manuscript. We thank Erwin de Blok who kindly provided the H{\sc i} surface density profiles. We are grateful to Nataliya Zaitseva for the support.
Development of the program package for analysis of $\chi^2$ maps has been supported by the project by Russian Foundation for Basic Research (RFBR) project No. 14-22-03006-ofi-m.
The tests of the effects of rotation curve truncation and angular resolution on the reliability of the parameters of dark halo were made with the support of RFBR grants No. 15-32-21062-a and 15-52-15050. 
The analysis and classification of $\chi^2$ maps and model rotation curves was performed with aid of the Russian President's grant No. MD-7355.2015.2. 
The surface photometry was done with the support of the Russian Science Foundation project 14-22-00041.

\bibliographystyle{mnras}
\bibliography{DM_project}

\appendix

   \setcounter{table}{0} \renewcommand{\thetable}{A.\arabic{table}}  

\begin{table*}
\begin{center}
\caption{The obtained parameters of the main components of the galaxies for Burkert dark halo density profile. 
The errors correspond to $1\sigma$ confidence limit. The columns contain the following data:
(1)~-- galaxy name;
(2) and (3)~-- radial scale and central density of the DM halo;
(4)~-- optical radius;
(5)~-- mass of DM halo inside of optical radius;
(6)~-- disc mass-to-light ratio at 3.6\ $\mu$m for all galaxies except for V-band in cases of \N925 and \N7331;
(7)~-- central surface density of bulge\label{parameters_burk};
(8)~-- reduced value of $\chi^2$;
\red{(9) ~-- the ratio between the dynamic mass  $M_{dyn}=R_{opt}v^2(R_{max})/G$ and the halo mass found from the rotation curve decomposition.}}
\renewcommand{\arraystretch}{1.5}
\begin{tabular}{lrlrl r rlrlrlrr}
\hline
NGC	&	\multicolumn{2}{c}{$R_s$}&	\multicolumn{2}{c}{$\rho_0$ }&	\multicolumn{1}{c}{$R_{opt}$}	&	\multicolumn{2}{c}{$M_{halo}$}	 &	\multicolumn{2}{c}{$M/L$} &	\multicolumn{2}{c}{$(I_0)_b$}&$\chi_r^2$&$\frac{M_{dyn}}{M_{halo}}$\\
&\multicolumn{2}{c}{kpc}&\multicolumn{2}{c}{$10^{-3}$ M$_{\odot}/$pc$^3$}& \multicolumn{1}{c}{kpc}&\multicolumn{2}{c}{$10^{10}$ M$_{\odot}$}&\multicolumn{2}{c}{M$_{\odot}/$L$_{\odot}$	}& \multicolumn{2}{c}{$10^{3}$ M$_{\odot}/$pc$^2$}&&\\
\hline
\hline
\multicolumn{14}{c}{Model A}\\
\hline
         925&      8.16& $^{+      2.03}_{-      1.40} $  &     17.23& $^{+      6.68}_{-      4.96} $  &     11.20&      2.66& $^{+      0.22}_{-      0.21} $  &      0.20& $^{+      0.20}_{-      0.20} $  &      \multicolumn{2}{c}{---}  &      0.71&1.28\\
        2403&      8.35& $^{+      0.86}_{-      0.85} $  &     19.13& $^{+      3.87}_{-      2.86} $  &      7.24&      1.33& $^{+      0.11}_{-      0.10} $  &      0.17& $^{+      0.02}_{-      0.02} $  &      0.60& $^{+      0.07}_{-      0.06} $  &      0.51&2.41\\
        2841&     12.89& $^{+      2.01}_{-      1.86} $  &     30.35& $^{+     11.72}_{-      7.58} $  &     13.67&     11.63& $^{+      1.83}_{-      1.51} $  &      0.91& $^{+      0.11}_{-      0.13} $  &    534.20& $^{+     34.69}_{-     35.04} $  &      0.24&2.27\\
        2903&      4.22& $^{+     19.57}_{-      0.84} $  &    148.06& $^{+    105.51}_{-    143.57} $  &     26.14&     16.78& $^{+      0.62}_{-      5.32} $  &      0.00& $^{+      0.40}_{-      0.00} $  &    336.87& $^{+     55.83}_{-    103.37} $  &      0.52&1.16\\
        2976&      2.14& $^{+      2.32}_{-      0.62} $  &    152.52& $^{+    123.59}_{-     85.47} $  &      3.80&      0.63& $^{+      0.11}_{-      0.10} $  &      0.16& $^{+      0.16}_{-      0.16} $  &     \multicolumn{2}{c}{---} &      0.88&1.03\\
        3031&      5.49& $^{+      1.19}_{-      1.04} $  &     74.35& $^{+     40.82}_{-     23.65} $  &      7.61&      3.55& $^{+      0.52}_{-      0.47} $  &      0.72& $^{+      0.09}_{-      0.09} $  &    233.74& $^{+     38.67}_{-     46.73} $  &      2.60&1.92\\
        3198&      4.40& $^{+      0.92}_{-      0.37} $  &     95.19& $^{+     24.02}_{-     37.16} $  &     13.38&      6.64& $^{+      0.39}_{-      0.82} $  &      0.09& $^{+      0.24}_{-      0.09} $  &      2.45& $^{+      0.86}_{-      0.86} $  &      0.45&1.04\\
        3521&      2.77& $^{+      2.37}_{-      0.50} $  &    467.90& $^{+    275.97}_{-    371.17} $  &      9.35&      9.09& $^{+      0.51}_{-      3.31} $  &      0.01& $^{+      0.27}_{-      0.01} $  &    288.27& $^{+     79.52}_{-     98.66} $  &      0.18&0.98\\
        3621&     24.43& $^{+      4.45}_{-      3.18} $  &      3.91& $^{+      0.64}_{-      0.58} $  &     10.44&      1.28& $^{+      0.12}_{-      0.12} $  &      0.70& $^{+      0.02}_{-      0.02} $  &     \multicolumn{2}{c}{---}  &      0.80&4.78\\
        4736&      1.72& $^{+      0.81}_{-      0.55} $  &    210.55& $^{+    386.62}_{-    131.15} $  &      5.30&      0.89& $^{+      0.25}_{-      0.22} $  &      0.35& $^{+      0.08}_{-      0.08} $  &     77.17& $^{+     28.30}_{-     35.20} $  &      1.12&1.83\\
        5055&     13.10& $^{+      1.56}_{-      1.58} $  &     12.64& $^{+      3.74}_{-      2.57} $  &     13.89&      5.08& $^{+      0.60}_{-      0.53} $  &      0.33& $^{+      0.02}_{-      0.03} $  &    402.21& $^{+     22.12}_{-     20.40} $  &      0.53&1.84\\
        6946&      3.92& $^{+      9.22}_{-      0.19} $  &    196.07& $^{+     26.62}_{-    179.48} $  &     11.88&      9.66& $^{+      0.26}_{-      4.75} $  &      0.01& $^{+      0.36}_{-      0.01} $  &     19.35& $^{+      1.63}_{-      5.54} $  &      0.83&1.13\\
        7331&    100.00& $^{+    \infty}_{-     45.63} $  &      2.43& $^{+      0.81}_{-      \infty} $  &     25.49&     13.65& $^{+      2.03}_{-      1.82} $  &      2.61& $^{+      0.32}_{-      0.36} $  &    767.44& $^{+     28.75}_{-     25.72} $  &      0.43&2.81\\
        7793&      2.89& $^{+      0.34}_{-      0.31} $  &     96.80& $^{+     24.22}_{-     19.37} $  &      5.24&      1.02& $^{+      0.07}_{-      0.06} $  &      0.30& $^{+      0.04}_{-      0.04} $  &      0.00& $^{+      0.00}_{-      0.00} $  &      2.32&1.23\\
\hline
\hline
\multicolumn{14}{c}{Model B}\\
\hline
         925&     11.63& $^{+      1.67}_{-      1.38} $  &     10.14& $^{+      1.41}_{-      1.32} $  &     11.20&      2.36& $^{+      0.12}_{-      0.12} $  &      0.54& $^{+      0.02}_{-      0.00} $  &       \multicolumn{2}{c}{---}  &      0.87&1.44\\
        2403&      9.13& $^{+      0.45}_{-      0.50} $  &     16.36& $^{+      1.58}_{-      1.07} $  &      7.24&      1.23& $^{+      0.06}_{-      0.03} $  &      0.60& $^{+      0.00}_{-      0.02} $  &      0.60& $^{+      0.00}_{-      0.00} $  &      0.61&2.61\\
        2841&      4.33& $^{+      0.07}_{-      0.07} $  &    364.20& $^{+     13.29}_{-     12.82} $  &     13.67&     25.17& $^{+      0.10}_{-      0.10} $  &      0.45& $^{+      0.00}_{-      0.00} $  &     33.85& $^{+      0.00}_{-      0.00} $  &      2.59&1.05\\
        2903&     83.05& $^{+     \infty}_{-     33.21} $  &      1.83& $^{+      0.44}_{-      \infty} $  &     26.14&     10.48& $^{+      0.49}_{-      0.52} $  &      0.60& $^{+      0.00}_{-      0.00} $  &     13.56& $^{+      0.00}_{-      0.00} $  &      1.91&1.86\\
        2976&    110.00& $^{+   \infty}_{-     98.20} $  &     29.50& $^{+      6.01}_{-     \infty} $  &      3.80&      0.66& $^{+      0.08}_{-      0.11} $  &      0.45& $^{+      0.00}_{-      0.00} $  &        \multicolumn{2}{c}{---}  &      1.64&0.99\\
        3031&      3.29& $^{+      0.22}_{-      0.21} $  &    220.01& $^{+     33.17}_{-     22.53} $  &      7.61&      4.72& $^{+      0.18}_{-      0.10} $  &      0.60& $^{+      0.00}_{-      0.02} $  &     46.85& $^{+      0.00}_{-      0.00} $  &      2.89&1.44\\
        3198&      6.14& $^{+      1.39}_{-      0.32} $  &     41.12& $^{+      4.98}_{-     15.40} $  &     13.38&      5.32& $^{+      0.12}_{-      0.66} $  &      0.45& $^{+      0.15}_{-      0.00} $  &      5.81& $^{+      0.00}_{-      0.00} $  &      0.77&1.3\\
        3521&      6.12& $^{+      1.81}_{-      0.57} $  &     62.31& $^{+     12.36}_{-     26.82} $  &      9.35&      4.83& $^{+      0.27}_{-      0.97} $  &      0.45& $^{+      0.06}_{-      0.00} $  &     49.42& $^{+      0.00}_{-      0.00} $  &      0.45&1.85\\
        3621&     15.49& $^{+      1.20}_{-      1.24} $  &      6.96& $^{+      0.82}_{-      0.65} $  &     10.44&      1.78& $^{+      0.09}_{-      0.08} $  &      0.60& $^{+      0.00}_{-      0.00} $  &       \multicolumn{2}{c}{---}  &      1.55&3.45\\
        4736&      2.87& $^{+      0.74}_{-      0.56} $  &     53.67& $^{+     28.02}_{-     18.41} $  &      5.30&      0.57& $^{+      0.05}_{-      0.06} $  &      0.45& $^{+      0.00}_{-      0.00} $  &     25.09& $^{+      0.00}_{-      0.00} $  &      1.24&2.88\\
        5055&     12.92& $^{+      1.85}_{-      1.46} $  &     12.93& $^{+      3.52}_{-      2.89} $  &     13.89&      5.12& $^{+      0.55}_{-      0.56} $  &      0.33& $^{+      0.02}_{-      0.02} $  &     88.34& $^{+      4.40}_{-      4.66} $  &      0.53&1.83\\
        6946&      8.84& $^{+      1.86}_{-      0.72} $  &     31.20& $^{+      3.76}_{-      8.29} $  &     11.88&      5.91& $^{+      0.10}_{-      0.48} $  &      0.30& $^{+      0.03}_{-      0.00} $  &     23.26& $^{+      0.00}_{-      0.00} $  &      0.86&1.85\\
        7331&      5.15& $^{+      0.59}_{-      0.61} $  &     90.34& $^{+     26.69}_{-     17.24} $  &     25.49&     15.71& $^{+      0.68}_{-      0.78} $  &      1.62& $^{+      0.13}_{-      0.00} $  &    546.76& $^{+      0.00}_{-      3.12} $  &      1.15&2.44\\
        7793&      5.11& $^{+      0.57}_{-      0.55} $  &     35.45& $^{+      4.91}_{-      4.31} $  &      5.24&      0.80& $^{+      0.03}_{-      0.04} $  &      0.45& $^{+      0.00}_{-      0.00} $  &     19.55& $^{+      0.00}_{-      0.00} $  &     10.20&1.58\\
\hline
\end{tabular}
\end{center}
\end{table*}

\begin{table*}
\begin{center}
\caption{The same parameters as in Tab.~\ref{parameters_burk}, but for pseudoisothermal dark halo density profile. \label{parameters_piso}}
\renewcommand{\arraystretch}{1.5}
\begin{tabular}{lrlrl r rlrlrlrr}
\hline
NGC	&	\multicolumn{2}{c}{$R_s$}&	\multicolumn{2}{c}{$\rho_0$ }&	\multicolumn{1}{c}{$R_{opt}$}	&	\multicolumn{2}{c}{$M_{halo}$}	 &	\multicolumn{2}{c}{$M/L$} &	\multicolumn{2}{c}{$(I_0)_b$}&$\chi_r^2$&$\frac{M_{dyn}}{M_{halo}}$\\
&\multicolumn{2}{c}{kpc}&\multicolumn{2}{c}{$10^{-3}$ M$_{\odot}/$pc$^3$}& \multicolumn{1}{c}{kpc}&\multicolumn{2}{c}{$10^{10}$ M$_{\odot}$}&\multicolumn{2}{c}{M$_{\odot}/$L$_{\odot}$	}& \multicolumn{2}{c}{$10^{3}$ M$_{\odot}/$pc$^2$}&&\\
\hline
\hline
\multicolumn{14}{c}{Model A}\\
\hline
        925&      5.60& $^{+      1.60}_{-      1.28} $  &     13.47& $^{+      6.83}_{-      3.90} $  &     11.20&      2.65& $^{+      0.24}_{-      0.21} $  &      0.28& $^{+      0.18}_{-      0.25} $  &     \multicolumn{2}{c}{---}  &      0.75&1.28\\
        2403&      3.95& $^{+      0.67}_{-      1.20} $  &     26.08& $^{+     23.34}_{-      5.79} $  &      7.24&      1.54& $^{+      0.30}_{-      0.12} $  &      0.14& $^{+      0.02}_{-      0.05} $  &      0.68& $^{+      0.11}_{-      0.05} $  &      0.48&2.09\\
        2841&      0.14& $^{+      3.56}_{-      \infty} $  &  71044.20& $^{+ \infty}_{-  70943.38} $  &     13.67&     22.00& $^{+      0.67}_{-      6.71} $  &      0.59& $^{+      0.15}_{-      0.04} $  &     46.87& $^{+    416.76}_{-      6.75} $  &      0.22&1.2\\
        2903&      0.12& $^{+      1.62}_{-      \infty} $  &  36571.68& $^{+  \infty}_{-  36394.87} $  &     26.14&     16.82& $^{+      0.94}_{-      1.39} $  &      0.14& $^{+      0.03}_{-      0.03} $  &    142.57& $^{+    152.29}_{-     12.52} $  &      0.41&1.16\\
        2976&      1.61& $^{+      1.23}_{-      0.57} $  &    104.00& $^{+    119.27}_{-     48.24} $  &      3.80&      0.65& $^{+      0.11}_{-      0.10} $  &      0.22& $^{+      0.11}_{-      0.21} $  &       \multicolumn{2}{c}{---}  &      0.88&1.01\\
        3031&      3.93& $^{+      1.24}_{-      1.10} $  &     48.57& $^{+     35.59}_{-     16.41} $  &      7.61&      3.11& $^{+      0.46}_{-      0.34} $  &      0.81& $^{+      0.06}_{-      0.08} $  &    181.63& $^{+     38.24}_{-     28.74} $  &      2.81&2.19\\
        3198&      2.65& $^{+      0.89}_{-      0.44} $  &     53.06& $^{+     23.14}_{-     22.84} $  &     13.38&      4.55& $^{+      0.27}_{-      0.41} $  &      0.61& $^{+      0.12}_{-      0.08} $  &      0.03& $^{+      0.73}_{-      0.03} $  &      0.55&1.52\\
        3521&      0.27& $^{+      2.28}_{-      0.01} $  &   8307.36& $^{+   1692.64}_{-   8210.40} $  &      9.35&      7.00& $^{+      0.64}_{-      2.22} $  &      0.28& $^{+      0.10}_{-      0.05} $  &      0.00& $^{+    245.94}_{-      0.00} $  &      0.42&1.28\\
        3621&     16.22& $^{+      2.60}_{-      2.50} $  &      3.32& $^{+      0.63}_{-      0.44} $  &     10.44&      1.28& $^{+      0.14}_{-      0.11} $  &      0.70& $^{+      0.01}_{-      0.02} $  &       \multicolumn{2}{c}{---}  &      0.79&4.79\\
        4736&      1.03& $^{+      1.02}_{-      \infty} $  &    130.20& $^{+ \infty}_{-     96.26} $  &      5.30&      0.67& $^{+      0.26}_{-      0.17} $  &      0.42& $^{+      0.06}_{-      0.05} $  &     42.17& $^{+     19.07}_{-     29.79} $  &      1.37&2.42\\
        5055&      0.34& $^{+      0.58}_{-      0.19} $  &   5007.36& $^{+  20632.33}_{-   4339.32} $  &     13.89&      9.68& $^{+      0.36}_{-      0.79} $  &      0.12& $^{+      0.02}_{-      0.02} $  &     89.86& $^{+    193.52}_{-     89.86} $  &      0.58&0.97\\
        6946&      3.71& $^{+      2.71}_{-      1.14} $  &     53.90& $^{+     57.77}_{-     32.77} $  &     11.88&      6.69& $^{+      1.10}_{-      1.24} $  &      0.25& $^{+      0.09}_{-      0.08} $  &     17.27& $^{+      1.43}_{-      1.86} $  &      0.80&1.63\\
        7331&      1.14& $^{+     \infty}_{-      0.64} $  &    656.70& $^{+   3031.20}_{-    \infty} $  &     25.49&     25.51& $^{+      3.10}_{-     12.28} $  &      0.00& $^{+      2.79}_{-      0.00} $  &    473.88& $^{+    298.80}_{-    188.03} $  &      0.37&1.5\\
        7793&      1.78& $^{+      0.31}_{-      0.25} $  &     84.14& $^{+     27.20}_{-     20.56} $  &      5.24&      1.01& $^{+      0.07}_{-      0.07} $  &      0.31& $^{+      0.04}_{-      0.05} $  &      0.00& $^{+      0.00}_{-      0.00} $  &      2.67&1.24\\
\hline
\hline
\multicolumn{14}{c}{Model B}\\
\hline
         925&      7.54& $^{+      1.13}_{-      0.89} $  &      8.82& $^{+      1.23}_{-      1.15} $  &     11.20&      2.41& $^{+      0.13}_{-      0.11} $  &      0.54& $^{+      0.03}_{-      0.00} $  &      \multicolumn{2}{c}{---}  &      0.85&1.41\\
        2403&      5.41& $^{+      0.31}_{-      0.66} $  &     15.56& $^{+      3.29}_{-      1.11} $  &      7.24&      1.27& $^{+      0.09}_{-      0.03} $  &      0.60& $^{+      0.00}_{-      0.03} $  &      0.60& $^{+      0.00}_{-      0.00} $  &      0.61&2.54\\
        2841&      0.22& $^{+      0.29}_{-     \infty} $  &  26556.87& $^{+ \infty}_{-  21512.11} $  &     13.67&     21.82& $^{+      0.84}_{-      0.62} $  &      0.59& $^{+      0.01}_{-      0.06} $  &     25.45& $^{+      8.39}_{-      0.07} $  &      0.22&1.21\\
        2903&     45.60& $^{+    \infty}_{-     14.85} $  &      1.70& $^{+      0.33}_{-     \infty} $  &     26.14&     10.68& $^{+      0.48}_{-      0.50} $  &      0.60& $^{+      0.00}_{-      0.00} $  &     13.56& $^{+      0.00}_{-      0.00} $  &      1.90&1.83\\
        2976&    100.00& $^{+  \infty}_{-     95.01} $  &     29.96& $^{+      5.01}_{-     \infty} $  &      3.80&      0.69& $^{+      0.08}_{-      0.17} $  &      0.45& $^{+      0.00}_{-      0.00} $  &      \multicolumn{2}{c}{---} &      1.63&0.95\\
        3031&      1.45& $^{+      0.09}_{-      0.50} $  &    314.56& $^{+    360.77}_{-     30.08} $  &      7.61&      4.66& $^{+      0.28}_{-      0.04} $  &      0.60& $^{+      0.00}_{-      0.02} $  &     46.85& $^{+      0.00}_{-     11.71} $  &      3.06&1.46\\
        3198&      2.88& $^{+      0.26}_{-      0.43} $  &     45.63& $^{+     16.65}_{-      6.31} $  &     13.38&      4.51& $^{+      0.23}_{-      0.09} $  &      0.60& $^{+      0.00}_{-      0.06} $  &      5.81& $^{+      0.00}_{-      0.00} $  &      0.88&1.54\\
        3521&      2.65& $^{+      1.07}_{-      0.45} $  &     87.46& $^{+     32.93}_{-     40.24} $  &      9.35&      4.56& $^{+      0.23}_{-      0.53} $  &      0.45& $^{+      0.04}_{-      0.00} $  &     49.42& $^{+      0.00}_{-      0.00} $  &      0.54&1.96\\
        3621&      9.23& $^{+      0.89}_{-      0.81} $  &      6.58& $^{+      0.85}_{-      0.72} $  &     10.44&      1.85& $^{+      0.09}_{-      0.09} $  &      0.60& $^{+      0.00}_{-      0.00} $  &      \multicolumn{2}{c}{---}  &      1.53&3.31\\
        4736&      1.64& $^{+      0.65}_{-      0.51} $  &     49.99& $^{+     45.17}_{-     21.70} $  &      5.30&      0.54& $^{+      0.05}_{-      0.06} $  &      0.45& $^{+      0.01}_{-      0.00} $  &     25.09& $^{+      0.00}_{-      0.00} $  &      1.38&3.01\\
        5055&      5.96& $^{+      1.89}_{-      0.83} $  &     17.46& $^{+      5.29}_{-      6.67} $  &     13.89&      5.41& $^{+      0.34}_{-      0.74} $  &      0.31& $^{+      0.04}_{-      0.01} $  &     89.75& $^{+      2.01}_{-      5.48} $  &      0.69&1.73\\
        6946&      5.31& $^{+      0.91}_{-      0.49} $  &     28.99& $^{+      3.89}_{-      6.14} $  &     11.88&      5.93& $^{+      0.14}_{-      0.27} $  &      0.30& $^{+      0.02}_{-      0.00} $  &     23.26& $^{+      0.00}_{-      0.00} $  &      0.83&1.84\\
        7331&      1.78& $^{+      0.42}_{-      1.10} $  &    190.95& $^{+   1056.05}_{-     62.02} $  &     25.49&     17.37& $^{+      0.59}_{-      0.87} $  &      1.62& $^{+      0.10}_{-      0.00} $  &    545.77& $^{+      0.98}_{-     79.59} $  &      0.69&2.21\\
        7793&      3.40& $^{+      0.35}_{-      0.35} $  &     29.95& $^{+      4.18}_{-      3.32} $  &      5.24&      0.81& $^{+      0.03}_{-      0.04} $  &      0.45& $^{+      0.00}_{-      0.00} $  &     19.55& $^{+      0.00}_{-      0.00} $  &     10.18&1.55\\
\hline
\end{tabular}
\end{center}
\end{table*}

\begin{table*}
\begin{center}
\caption{The same parameters as in Tab.~\ref{parameters_burk}, but for NFW dark halo density profile \label{parameters_nfw}}
\renewcommand{\arraystretch}{1.5}
\begin{tabular}{lrlrl r rlrlrlrr}
\hline
NGC	&	\multicolumn{2}{c}{$R_s$}&	\multicolumn{2}{c}{$\rho_0$ }&	\multicolumn{1}{c}{$R_{opt}$}	&	\multicolumn{2}{c}{$M_{halo}$}	 &	\multicolumn{2}{c}{$M/L$} &	\multicolumn{2}{c}{$(I_0)_b$}&$\chi_r^2$&$\frac{M_{dyn}}{M_{halo}}$\\
&\multicolumn{2}{c}{kpc}&\multicolumn{2}{c}{$10^{-3}$ M$_{\odot}/$pc$^3$}& \multicolumn{1}{c}{kpc}&\multicolumn{2}{c}{$10^{10}$ M$_{\odot}$}&\multicolumn{2}{c}{M$_{\odot}/$L$_{\odot}$	}& \multicolumn{2}{c}{$10^{3}$ M$_{\odot}/$pc$^2$}&&\\
\hline
\hline
\multicolumn{14}{c}{Model A}\\
\hline
        925&    100.00& $^{+      \infty}_{-     12.81} $  &      0.41& $^{+      0.07}_{-     \infty} $  &     11.20&      2.79& $^{+      0.08}_{-      0.09} $  &      0.00& $^{+      0.00}_{-      0.00} $  &       \multicolumn{2}{c}{---} &      1.41&1.22\\
        2403&     17.43& $^{+      5.87}_{-      3.99} $  &      5.02& $^{+      2.93}_{-      1.93} $  &      7.24&      1.80& $^{+      0.16}_{-      0.16} $  &      0.11& $^{+      0.02}_{-      0.02} $  &      0.41& $^{+      0.02}_{-      0.03} $  &      0.51&1.79\\
        2841&     17.03& $^{+      5.80}_{-      4.81} $  &     18.98& $^{+     19.92}_{-      8.61} $  &     13.67&     16.95& $^{+      2.90}_{-      2.27} $  &      0.67& $^{+      0.14}_{-      0.16} $  &    414.64& $^{+     11.48}_{-     26.65} $  &      0.20&1.56\\
        2903&      4.69& $^{+      2.15}_{-     \infty} $  &    137.03& $^{+   \infty}_{-     82.32} $  &     26.14&     18.39& $^{+      0.69}_{-      1.37} $  &      0.00& $^{+      0.05}_{-      0.00} $  &    184.36& $^{+     82.86}_{-     75.98} $  &      0.42&1.06\\
        2976&    100.00& $^{+     \infty}_{-     42.36} $  &      0.99& $^{+      0.75}_{-     \infty} $  &      3.80&      0.85& $^{+      0.03}_{-      0.07} $  &      0.00& $^{+      0.04}_{-      0.00} $  &      0.00& $^{+      0.00}_{-      0.00} $  &      2.44&0.77\\
        3031&      9.49& $^{+      6.95}_{-      3.08} $  &     27.24& $^{+     33.74}_{-     16.97} $  &      7.61&      4.20& $^{+      0.65}_{-      0.64} $  &      0.69& $^{+      0.10}_{-      0.09} $  &    137.73& $^{+     27.98}_{-     27.83} $  &      2.87&1.62\\
        3198&     15.77& $^{+      4.28}_{-      2.60} $  &      6.11& $^{+      2.84}_{-      2.34} $  &     13.38&      4.67& $^{+      0.39}_{-      0.48} $  &      0.57& $^{+      0.13}_{-      0.10} $  &      \multicolumn{2}{c}{---}  &      0.73&1.48\\
        3521&      2.85& $^{+      3.05}_{-      0.29} $  &    522.54& $^{+    146.58}_{-    438.17} $  &      9.35&     10.48& $^{+      0.35}_{-      3.15} $  &      0.01& $^{+      0.20}_{-      0.01} $  &      0.00& $^{+    172.84}_{-      0.00} $  &      0.19&0.85\\
        3621&    100.00& $^{+     \infty}_{-     20.35} $  &      0.34& $^{+      0.12}_{-      \infty} $  &     10.44&      2.06& $^{+      0.07}_{-      0.06} $  &      0.56& $^{+      0.01}_{-      0.01} $  &      \multicolumn{2}{c}{---}  &      0.79&2.98\\
        4736&      1.89& $^{+      2.12}_{-      0.91} $  &    205.32& $^{+    956.12}_{-    170.88} $  &      5.30&      1.04& $^{+      0.33}_{-      0.30} $  &      0.33& $^{+      0.09}_{-      0.09} $  &     63.77& $^{+     22.90}_{-     27.94} $  &      1.24&1.57\\
        5055&     15.27& $^{+      6.38}_{-      5.64} $  &     10.19& $^{+     18.17}_{-      5.17} $  &     13.89&      7.77& $^{+      1.86}_{-      1.07} $  &      0.20& $^{+      0.05}_{-      0.09} $  &    384.54& $^{+      7.70}_{-     11.22} $  &      0.56&1.21\\
        6946&     29.76& $^{+     59.55}_{-     15.61} $  &      3.97& $^{+     10.56}_{-      3.09} $  &     11.88&      6.64& $^{+      1.27}_{-      0.87} $  &      0.26& $^{+      0.06}_{-      0.09} $  &     10.35& $^{+      1.01}_{-      2.15} $  &      0.89&1.64\\
        7331&     11.97& $^{+     \infty}_{-      4.65} $  &     23.69& $^{+     45.34}_{-    \infty} $  &     25.49&     23.47& $^{+      1.62}_{-      5.96} $  &      0.00& $^{+      1.40}_{-      0.00} $  &    599.59& $^{+    162.56}_{-    159.26} $  &      0.46&1.63\\
        7793&      9.59& $^{+      3.21}_{-      1.99} $  &     13.56& $^{+      7.02}_{-      5.16} $  &      5.24&      1.24& $^{+      0.10}_{-      0.10} $  &      0.10& $^{+      0.07}_{-      0.07} $  &      0.00& $^{+      0.00}_{-      0.00} $  &      3.51&1.01\\
\hline
\hline
\multicolumn{14}{c}{Model B}\\
\hline
        925&    100.00& $^{+     \infty}_{-      7.90} $  &      0.30& $^{+      0.03}_{-      \infty} $  &     11.20&      2.04& $^{+      0.09}_{-      0.09} $  &      0.54& $^{+      0.00}_{-      0.00} $  &    \multicolumn{2}{c}{---} &      2.96&1.67\\
        2403&     27.83& $^{+      3.87}_{-      2.94} $  &      2.31& $^{+      0.40}_{-      0.39} $  &      7.24&      1.55& $^{+      0.03}_{-      0.04} $  &      0.45& $^{+      0.01}_{-      0.00} $  &      0.60& $^{+      0.00}_{-      0.06} $  &      0.57&2.07\\
        2841&      7.30& $^{+      0.23}_{-      0.45} $  &    125.70& $^{+     17.05}_{-      8.81} $  &     13.67&     24.74& $^{+      0.28}_{-      0.28} $  &      0.45& $^{+      0.02}_{-      0.00} $  &     33.85& $^{+      0.00}_{-      0.68} $  &      0.85&1.07\\
        2903&      6.94& $^{+      1.24}_{-      0.20} $  &     42.74& $^{+      2.02}_{-     11.79} $  &     26.14&     13.85& $^{+      0.45}_{-      0.23} $  &      0.30& $^{+      0.01}_{-      0.00} $  &     13.56& $^{+      0.00}_{-      1.61} $  &      2.16&1.41\\
        2976&    100.00& $^{+    \infty}_{-     57.26} $  &      0.29& $^{+      0.41}_{-     \infty} $  &      3.80&      0.25& $^{+      0.03}_{-      0.03} $  &      0.45& $^{+      0.01}_{-      0.00} $  &      \multicolumn{2}{c}{---} &      4.68&2.66\\
        3031&      6.69& $^{+      1.83}_{-      1.23} $  &     56.36& $^{+     29.25}_{-     19.92} $  &      7.61&      4.82& $^{+      0.31}_{-      0.14} $  &      0.60& $^{+      0.00}_{-      0.04} $  &     37.45& $^{+      9.40}_{-      2.32} $  &      2.91&1.41\\
        3198&     17.46& $^{+      1.78}_{-      2.96} $  &      4.98& $^{+      2.35}_{-      0.78} $  &     13.38&      4.49& $^{+      0.38}_{-      0.11} $  &      0.60& $^{+      0.00}_{-      0.10} $  &      5.81& $^{+      0.00}_{-      0.00} $  &      1.34&1.54\\
        3521&     15.41& $^{+      3.87}_{-      2.95} $  &     10.27& $^{+      4.23}_{-      3.38} $  &      9.35&      4.56& $^{+      0.17}_{-      0.36} $  &      0.45& $^{+      0.02}_{-      0.00} $  &     49.42& $^{+      0.00}_{-      0.00} $  &      0.52&1.96\\
        3621&    100.00& $^{+     \infty}_{-     20.25} $  &      0.34& $^{+      0.12}_{-      \infty} $  &     10.44&      2.05& $^{+      0.08}_{-      0.05} $  &      0.56& $^{+      0.01}_{-      0.02} $  &      \multicolumn{2}{c}{---}  &      0.79&2.99\\
        4736&      7.80& $^{+      6.18}_{-      3.05} $  &      8.05& $^{+     10.82}_{-      4.86} $  &      5.30&      0.55& $^{+      0.04}_{-      0.06} $  &      0.45& $^{+      0.00}_{-      0.00} $  &     25.09& $^{+      0.00}_{-      0.00} $  &      1.47&2.99\\
        5055&     31.76& $^{+      4.01}_{-      3.56} $  &      2.40& $^{+      0.51}_{-      0.42} $  &     13.89&      5.65& $^{+      0.19}_{-      0.20} $  &      0.30& $^{+      0.00}_{-      0.00} $  &     77.15& $^{+      2.21}_{-      1.02} $  &      0.68&1.66\\
        6946&    100.00& $^{+      \infty}_{-     21.59} $  &      0.77& $^{+      0.24}_{-     \infty} $  &     11.88&      5.84& $^{+      0.07}_{-      0.16} $  &      0.30& $^{+      0.00}_{-      0.00} $  &     23.26& $^{+      0.00}_{-      0.00} $  &      1.21&1.87\\
        7331&     12.46& $^{+      3.22}_{-      2.29} $  &     16.02& $^{+      7.67}_{-      5.56} $  &     25.49&     17.19& $^{+      0.90}_{-      0.92} $  &      1.62& $^{+      0.13}_{-      0.00} $  &    546.76& $^{+      0.00}_{-      3.75} $  &      0.79&2.23\\
        7793&    100.00& $^{+     \infty}_{-     16.29} $  &      0.44& $^{+      0.10}_{-      \infty} $  &      5.24&      0.70& $^{+      0.03}_{-      0.03} $  &      0.45& $^{+      0.00}_{-      0.00} $  &     19.55& $^{+      0.00}_{-      0.00} $  &     14.61&1.79\\
\hline
\end{tabular}
\end{center}
\end{table*}

\label{lastpage}

\end{document}